\documentclass[12pt]{article}
\usepackage{epsfig,amsfonts,amsmath,amssymb,amscd,float}
\usepackage{graphicx,caption,float,subcaption}
\usepackage{booktabs,multirow,siunitx}
\usepackage{authblk}
\usepackage{cite}
\usepackage{booktabs}
\usepackage{multirow}
\usepackage{soul}
\usepackage{microtype}
\usepackage{epstopdf}
\usepackage{graphicx}
\usepackage{ifthen}
\usepackage{xurl}
\usepackage{algorithm}
\usepackage{algorithmic}
\usepackage{xcolor}
\usepackage{amsthm}
\usepackage{cite}

\newtheorem{theorem}{Theorem}[section]

\newtheorem{proposition}[theorem]{Proposition}

\theoremstyle{definition}

\theoremstyle{remark}

\numberwithin{equation}{section}
\makeatletter
\newcommand\stroke[1]{\mathpalette\stroke@aux{#1}}
\def\stroke@aux#1#2{%
  \ooalign{%
    \hfil$#1-$\hfil\cr
    \hfil$#1#2$\hfil\cr
  }%
}
\makeatother

\begin{document}

\title{Syringe reuse in HCV transmission among people who inject drugs: behavioral thresholds and optimal control}

\author[1]{Mohammad Zaman\thanks{imzaman82@gmail.com}}
\author[2]{Emel Khan\thanks{emalkhan@kust.edu.pk(corresponding)}}
\author[3]{Ishfaq Ahmad\thanks{ishfaqmaths899@gmail.com}}
\author[4]{Hiromi Seno\thanks{seno.math.is@tohoku.ac.jp}}

\affil[1]{Department of Mathematics, Government Post Graduate College, Charsadda, 24420, Khyber Pakhtunkhwa, Pakistan. }
\affil[2]{Institute of Numerical Sciences, Kohat University of Science and Technology, Kohat, 26000, Khyber Pakhtunkhwa, Pakistan. }
\affil[3]{Knight Cancer Institute, Oregon Health and Science University, Portland, 97239, Oregon, USA.}
\affil[4]{
Graduate School of Information Sciences, Tohoku University, Sendai, 980-8579, Miyagi, Japan.}

\date{}
\maketitle

\setcounter{page}{1}

\abstract{Hepatitis C virus (HCV) transmission among people who inject drugs (PWID) is driven largely by syringe sharing and reuse, yet existing models rarely treat reuse as a behavior that individuals adopt and abandon. In this work, we develop and analyze a four-compartment epidemic model that explicitly incorporates syringe reuse as a distinct behavioral process, introducing a high-risk susceptible subgroup that acquires infection through both direct contact and syringe-mediated transmission. We establish the well-posedness and biological feasibility of the model, derive the basic reproduction number, and analyze the existence and stability of the disease-free and endemic equilibria along with closed-form behavioral threshold conditions for syringe-reuse adoption and cessation. Global sensitivity and scenario analyses reveal that while transmission and recovery parameters dominate epidemic potential, behavioral parameters governing the adoption and cessation of syringe reuse play a crucial role in disease persistence and control thresholds. The model's projected infection dynamics are broadly consistent with observed U.S. HCV incidence trends. We further formulate an optimal control problem targeting reuse initiation and cessation, and show through cost-effectiveness analysis that combined behavioral interventions outperform single-component strategies in reducing both infection and high-risk behavior. These results indicate that behavioral prevention aimed at syringe reuse is a necessary component of sustained HCV control among PWID.}


\section{Introduction}

Drug addiction and the spread of infectious diseases are closely interconnected, posing significant health risks for people who inject drugs (PWID) and the wider community. The reuse of contaminated syringes has long been recognized as a major driver of these risks, and injection equipment use has increased in both developed and developing countries \cite{hayashi2018declining}. In the United States (US), an estimated 6.5 million individuals have injected drugs at some point in their lives, with approximately 1 million reporting recent injection drug use \cite{Lansky2014}. Reusing syringes or other injection equipment facilitates transmission of blood-borne infections such as HIV, hepatitis B, HCV, and various bacterial diseases \cite{Stark1997, Binka2015, Scheibe2019, elaiw2024global}, and most HCV infections in the US occur through sharing or reusing contaminated syringes \cite{thorpe2002risk, paquette2018injection}. The World Health Organization has set elimination targets of an 80\% reduction in HCV incidence by 2030, making a quantitative understanding of transmission dynamics among PWID especially timely \cite{Martin2018review}.

Mathematical modeling of epidemic dynamics is fundamental for understanding how infectious diseases spread and for deriving key epidemiological indicators such as the basic reproduction number \cite{Kermack1927, Martcheva2015}, and behavioral and awareness-based strategies are known to play a crucial role in reducing risky injecting practices alongside biomedical interventions \cite{sacks2012behavioural, hagan2013best}. Mathematical models of blood-borne infection among PWID have accordingly examined syringe reuse and sharing from several angles \cite{kaplan1992model, homer1991model}: transmission dynamics and typical injecting behavior \cite{murray2003impact, hutchinson2006modelling}, the impact of needle-syringe and opioid-substitution programs \cite{vickerman2006model, vickerman2007modelling, kwon2009impact}, antiviral treatment strategies \cite{martin2011mathematical}, and extensions incorporating partial immunity, reinfection, or network heterogeneity in sharing behavior \cite{elbasha2013model, kretzschmar2004modelling, greenhalgh1997mathematical}. More recent work has moved toward elimination contexts, showing that combined antiviral scale-up, screening, and harm-reduction strategies can drive $\mathcal{R}_0$ below the epidemic threshold \cite{ScottIceland2018, Marquez2020, Scott2016}, while related models of addiction and behavior more broadly including opiate-addiction dynamics \cite{White2007}, education/treatment-stratified PWID models \cite{Tresna2022}, and network-based sharing models \cite{Bellerose2021, Gicquelais2023} show that structural and behavioral interventions substantially shape transmission.

Despite this breadth, key behavioral mechanisms associated with syringe reuse remain insufficiently captured in standard epidemic frameworks. Injecting risk is typically incorporated indirectly through modified transmission coefficients or auxiliary awareness classes, rather than through an explicit susceptible behavioral subgroup with its own transition dynamics. This limits the ability of such models to disentangle the epidemiological impact of syringe reuse behavior from that of infection alone: no prior mean-field compartmental model has simultaneously represented the behavioral adoption of syringe reuse, the dual (direct and indirect) transmission pathways reusers face, and the dynamic cessation of that behavior as distinct, analytically tractable processes.

In this work, we address this gap by developing a four-compartment epidemic model that explicitly incorporates syringe reuse as a behavioral process distinct from infection. The syringe reuser compartment ($U$) represents a high-risk subgroup of susceptible individuals who repeatedly reuse previously used syringes, acquiring infection through direct contact with infectious individuals and through indirect, syringe-mediated transmission proportional to infectious prevalence. The model also incorporates behavioral cessation, allowing reusers to abandon risky practices and return to the susceptible class. This formulation enables a systematic investigation of how syringe reuse behavior interacts with infection dynamics, modifies epidemic thresholds, and influences the effectiveness of behavioral and biomedical interventions.

\section{Assumptions and modeling}
\label{sec2}

\begin{figure}[H]
    \centering  
 \includegraphics[width=0.7\textwidth]
 {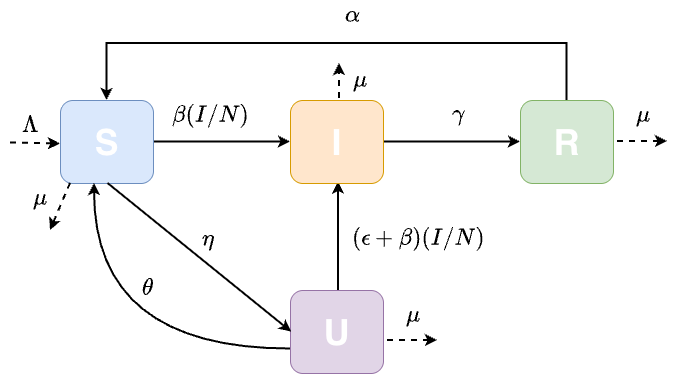}
     \caption{Schematic diagram of proposed model.}
    \label{fig:schematic}
\end{figure}

We consider an epidemic system in which the total population is divided into four epidemiological compartments: susceptible individuals ($S$), syringe reusers ($U$), infectious individuals ($I$), and recovered individuals ($R$), as illustrated in Figure \ref{fig:schematic}.

The compartment $U$ represents a high risk subgroup of susceptible individuals who repeatedly reuse previously used syringes. The model is based on the following assumptions:

\begin{enumerate} 
\item All newly recruited individuals (through birth) enter the susceptible compartment at a constant net recruitment rate.

\item Disease induced mortality is negligible; all individuals experience natural mortality at a constant per capita rate.

\item Infectious individuals recover at a constant rate and move to the recovered compartment. 

\item Recovered individuals experience waning immunity and may return to the susceptible compartment.

\item Susceptible individuals may adopt syringe reuse behavior and move from $S$ to the syringe reuser compartment 
$U$. Individuals in 
$U$ remain epidemiologically susceptible until infection occurs.

\item 
Infection is transmitted directly through effective contact with infectious individuals. The per capita force of infection is
$\lambda(t) = \beta I(t)/N(t),$ where $\beta$ is the transmission coefficient for direct contact. The net rates of direct infection are $\beta (I/N) S$  for susceptibles and $\beta (I/N) U$ for syringe reusers.

\item In addition to direct transmission, syringe reusers may acquire infection indirectly by  using a previously used and contaminated syringe. Following a mean field assumption treating the previous user of a reused syringe as a randomly selected individual from the population the probability that a syringe is contaminated is proportional to the infective fraction $I/N$.  Under this assumption, indirect infection among reusers occurs at net rate 
$\varepsilon(I/N) \, U$ where $\varepsilon$ is the infection coefficient for indirect transmission via syringe reuse.

\item Syringe reusers may cease risky injection behavior at rate $\theta$ and return to the susceptible compartment $S$. This cessation represents a behavioral change and does not confer any epidemiological immunity to infection.
\end{enumerate}

Under these assumptions, the total population size $N(t)=S(t)+U(t)+I(t)+R(t)$ satisfies the demographic equation
\[
\frac{dN}{dt}=\Lambda-\mu N,
\]
whose solution is
\[
N(t)=\frac{\Lambda}{\mu}+\left(N(0)-\frac{\Lambda}{\mu}\right)e^{-\mu t}.
\]
Hence $N(t)\to \Lambda/\mu$ as $t\to\infty$. In numerical simulations on an absolute population scale, we choose $\Lambda=\mu N(0)$ so that the total population remains constant near its initial level.

 The assumptions lead to the following $SIR+U$ model:
\begin{align}
\begin{array}{rl}
\dfrac{dS}{dt} &= \Lambda - \beta \dfrac{I}{N} S - \eta S + \alpha R + \theta U - \mu S,\\[3mm]
\dfrac{dU}{dt} &= \eta S - \beta \dfrac{I}{N} U - \varepsilon \dfrac{I}{N} U - \theta U
                 - \mu U, \\[3mm]
\dfrac{dI}{dt} &= \beta \dfrac{I}{N} S + \beta \dfrac{I}{N} U + \varepsilon \dfrac{I}{N} U  - \gamma I - \mu I,
                 \\[3mm]
\dfrac{dR}{dt} &= \gamma I  - \alpha R - \mu R.
\end{array}
\label{SIRU_model}
\end{align}
 with initial conditions $(S(0), U(0), I(0), R(0)) = (S_0, U_0, I_0, 0)$. 
 
 All rate parameters in system~\eqref{SIRU_model} correspond directly to the transitions described in the above assumptions; all parameters are positive and summarized, with baseline values and sources, in Table~\ref{table:params}.

\begin{table}[H]
\caption{Model parameters, descriptions, baseline values, and sources.
\label{table:params}}
\begin{tabular*}{\columnwidth}{@{\extracolsep\fill}l p{0.55\columnwidth} c p{0.22\columnwidth}@{}}
\toprule
\textbf{Parameter} & \textbf{Description} & \textbf{Value} & \textbf{Source} \\
\midrule
$\Lambda$ & Net recruitment rate into the population & $\mu N$ & assumed \\
$\mu$ & Natural mortality rate, uniform across compartments & 0.009 & \cite{naturaldeathrate} \\
$\beta$ & Direct transmission coefficient for infection via contact with infectious individuals & 0.638 & fitted \\
$\varepsilon$ & Indirect syringe-mediated transmission coefficient for reusers & 0.27 & \cite{leyva2020} \\
$\gamma$ & Recovery rate of infectious individuals & 0.5 & \cite{elbasha2013model} \\
$\eta$ & Rate of adopting syringe reuse behavior & 0.00165 & \cite{bradley2023estimated} \\
$\theta$ & Rate of cessation of syringe reuse & 0.056 & \cite{stocks2020} \\
$\alpha$ & Waning immunity rate & 0.025 & \cite{elbasha2013model} \\
\bottomrule
\end{tabular*}
\vspace{0.2cm}
\end{table}

\section{Well-posedness and biological feasibility}

In this section, we establish the mathematical well-posedness of the proposed model by
demonstrating positivity, boundedness, and global existence and uniqueness of solutions.
These properties guarantee that the model is biologically meaningful and mathematically
sound for epidemiological analysis.

\subsection{Invariant region and boundedness}

Since the state variables $S(t)$, $U(t)$, $I(t)$, and $R(t)$ represent population sizes, they
must remain nonnegative and bounded for all $t \ge 0$. We define the biologically feasible
region as
\[
\Omega = \left\{ (S,U,I,R) \in \mathbb{R}_+^4 \; : \; S+U+I+R \le \frac{\Lambda}{\mu} \right\}.
\]

\begin{theorem}[Boundedness and positive invariance]
\label{thm:boundedness}
For any nonnegative initial condition
$(S(0),U(0),I(0),R(0)) \in \Omega$, the region $\Omega$ is positively invariant and all
solutions of system~\eqref{SIRU_model} remain bounded for all $t \ge 0$.
\end{theorem}

\begin{proof}
Summing the equations in system~\eqref{SIRU_model} and solving yields
\[
N(t) = \frac{\Lambda}{\mu} + \left( N(0) - \frac{\Lambda}{\mu} \right) e^{-\mu t}.
\]
Hence,
\[
\lim_{t \to \infty} N(t) = \frac{\Lambda}{\mu},
\]
and
\[
0 \le N(t) \le \max\!\left\{N(0),\frac{\Lambda}{\mu}\right\}, \qquad \forall \,t \ge 0.
\]
Therefore, solutions starting in $\Omega$ remain in $\Omega$ for all $t \ge 0$, and the
region is positively invariant. Consequently, all solutions are bounded.
\end{proof}

\subsection{Positivity of solutions}

\begin{theorem}[Positivity of solutions]
\label{thm:positivity}
If $(S(0),U(0),I(0),R(0)) \in \mathbb{R}_+^4$, then the solution
$(S(t),U(t),I(t),R(t))$ of system~\eqref{SIRU_model} satisfies
\[
S(t)\ge0,\quad U(t)\ge0,\quad I(t)\ge0,\quad R(t)\ge0,
\qquad \forall t \ge 0.
\]
\end{theorem}

\begin{proof}
We argue by contradiction. Suppose that there exists a first time $t_1>0$ at which one of
the state variables becomes zero while the others remain nonnegative on $[0,t_1)$.

If $I(t_1)=0$, note that the third equation of \eqref{SIRU_model} can be
written as
\[
\frac{dI}{dt}=I\,g(S,U,I,R),
\]
where $g$ is continuous on $\Omega$. Therefore,
\[
I(t)=I(0)\exp\!\left(\int_0^t g(\tau)\,d\tau\right)\ge 0,\qquad t\ge 0,
\]
so $I(t)$ cannot become negative.

A similar argument holds for the other state variables: evaluating the
remaining equations of \eqref{SIRU_model} at their respective zeros gives
\[
\dot S\big|_{S=0}=\Lambda+\alpha R+\theta U>0,\qquad
\dot U\big|_{U=0}=\eta S\ge 0,\qquad
\dot R\big|_{R=0}=\gamma I\ge 0,
\]
each contradicting the assumption that the variable decreases below zero.
Therefore none of the state variables can become negative, and all remain
nonnegative for all $t\ge 0$.
\end{proof}

\subsection{Existence and uniqueness of solutions}

\begin{theorem}[Global existence and uniqueness]
\label{thm:existence_uniqueness}
For any initial condition
$(S(0),U(0),I(0),R(0)) \in \mathbb{R}_+^4$, system~\eqref{SIRU_model} admits a unique global
solution
\[
(S(t),U(t),I(t),R(t)) \in C^1\!\left([0,\infty),\mathbb{R}^4\right).
\]
\end{theorem}

\begin{proof}
The right-hand side of system~\eqref{SIRU_model} consists of continuously differentiable
functions on $\mathbb{R}_+^4$, and hence is locally Lipschitz continuous. By the
Picard--Lindel\"of theorem, a unique local solution exists.

By Theorems~\ref{thm:boundedness} and~\ref{thm:positivity}, the solution remains in the
compact set $\Omega$ for all $t\ge0$. Therefore, no finite-time blow-up is possible, and the
local solution can be extended globally. Uniqueness follows directly from the Lipschitz property.
\end{proof}

\noindent
Together, Theorems~\ref{thm:boundedness}--\ref{thm:existence_uniqueness} establish that
system~\eqref{SIRU_model} is mathematically well-posed and biologically feasible. All model
solutions remain nonnegative, bounded, and uniquely determined for all time, ensuring that
the model provides a reliable framework for analytical and numerical investigation of the
epidemic dynamics.

\section{Disease free equilibrium and basic reproduction number}
\label{subsec:DFE_R0}

\subsection{Disease free equilibrium (DFE)}
\label{subsec:DFE}

The DFE corresponds to the steady state at which
infection is absent from the population. For the present model, this occurs when $I = R = 0.$ Since syringe reuse is a behavioral process that may persist independently of
infection, the syringe reuser compartment may remain nonzero at the DFE.
Accordingly, the disease-free equilibrium is denoted by
\[
E_0 = (S^0,\, U^0,\, 0,\, 0).
\]

Solving the steady-state equations of the uninfected subsystem yields
\begin{equation}
E_0 =
\left(
\dfrac{\Lambda(\theta+\mu)}{\mu(\theta+\eta+\mu)},\;
\dfrac{\eta\Lambda}{\mu(\theta+\eta+\mu)},\;
0,\;
0
\right).
\label{E0}
\end{equation}
The total population at the DFE satisfies $N^0 = S^0 + U^0 = \Lambda/\mu$,
which is consistent with demographic balance.
\medskip

\subsection{Basic reproduction number}
\label{subsec:R0}

The basic reproduction number, denoted by $\mathcal{R}_0$, is defined as the
expected number of secondary infections generated by a typical infectious
individual introduced into a wholly susceptible population \cite{Diekmann1990, vandenDriessche2002}.

We compute $\mathcal{R}_0$ using the next-generation matrix method
\cite{Diekmann1990, vandenDriessche2002}. Since the model contains a single
infected compartment, $\mathcal{R}_0$ reduces to the spectral radius of a
scalar next-generation matrix. Straightforward calculations
(Appendix~\ref{appendix_R0}) yield

\begin{equation}
\mathcal{R}_0 =
\frac{\beta}{\gamma+\mu} +
\frac{\varepsilon\,\eta}{(\eta+\theta+\mu)(\gamma+\mu)}.
\label{R0}
\end{equation}

Expression~\eqref{R0} shows that the transmission potential consists of two
additive components. The first term corresponds to direct transmission through
effective contact, while the second term represents syringe-mediated
transmission among reusers. The latter contribution is governed by the rate of
entry into syringe reuse ($\eta$) and the average duration spent in the reuse
compartment $(\eta+\theta+\mu)^{-1}$. The basic reproduction number depends \emph{linearly}
on the direct transmission coefficient $\beta$ and the syringe-mediated
transmission coefficient $\varepsilon$. As a result, proportional changes in
either parameter lead to proportional changes in $\mathcal{R}_0$, which is
consistent with the linear trends observed in the scenario analysis.

In contrast, $\mathcal{R}_0$ depends \emph{nonlinearly} on the syringe reuse
adoption rate $\eta$ through the saturating term
\[
\frac{\varepsilon\,\eta}{\eta+\theta+\mu}.
\]
This term is monotonically increasing in $\eta$ but bounded above, reflecting
the fact that syringe-mediated transmission cannot increase indefinitely with
behavioral adoption. In particular,
\[
\lim_{\eta \to \infty} \mathcal{R}_0
=
\frac{\beta + \varepsilon}{\gamma + \mu},
\]
which provides an explicit upper bound on epidemic potential determined by the
combined transmission intensity and the average duration of infectiousness.

Biologically, this implies that while reducing syringe reuse adoption can
substantially lower transmission, increases in adoption beyond a certain level
yield diminishing returns in epidemic growth. This highlights the importance of
combining behavioral interventions with reductions in per-contact transmission
and infectious duration.

 \subsubsection{Global Sensitivity Analysis}

\begin{figure}[H]
\centering
\includegraphics[width=0.7\textwidth]{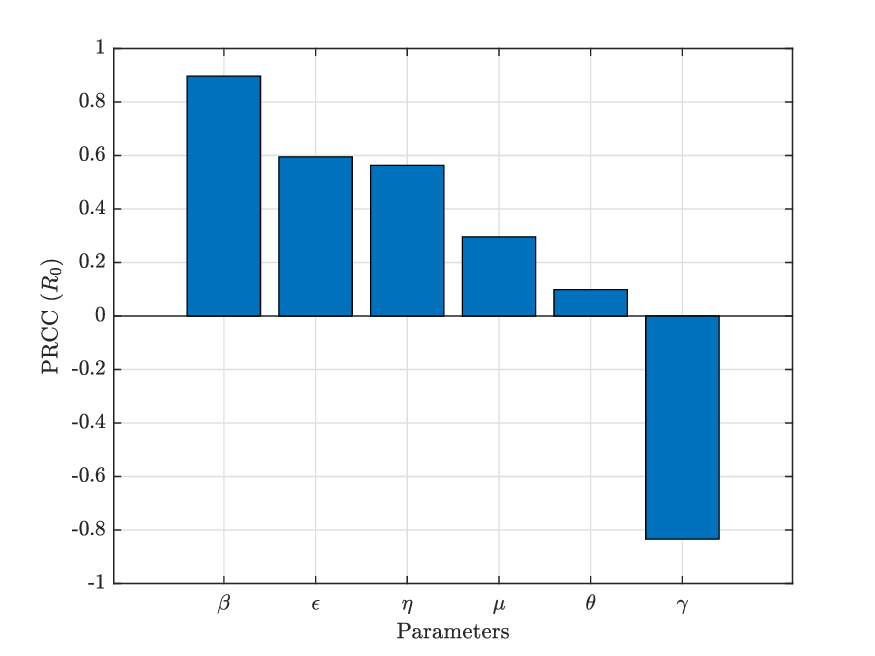}
\caption{Partial Rank Correlation Coefficients (PRCC) of $\mathcal{R}_0$.}
\label{fig:PRCC_R0}
\end{figure}

To rank the parameters influencing disease transmission, we performed a global sensitivity analysis of $\mathcal{R}_0$ using Latin Hypercube Sampling with Partial Rank Correlation Coefficients (PRCC) \cite{marino2008methodology}, varying all parameters in \eqref{R0} simultaneously across 1000 LHS-generated parameter sets; a positive PRCC indicates that increasing the parameter increases $\mathcal{R}_0$, and vice versa. Figure~\ref{fig:PRCC_R0} shows that the direct transmission coefficient $\beta$ has the highest positive correlation with $\mathcal{R}_0$, followed by the syringe-related coefficient $\varepsilon$ and the reuse adoption rate $\eta$ underscoring the role of syringe misuse alongside direct contact in sustaining transmission. Natural mortality $\mu$ and cessation rate $\theta$ are comparatively weak drivers, while the recovery rate $\gamma$ has the largest negative PRCC, confirming that improving treatment and recovery is a highly effective lever for suppressing spread alongside reductions in direct and syringe-mediated transmission.

\subsubsection{Scenario analysis of $\mathcal{R}_0$}

\begin{figure}[H]
\centering
\includegraphics[width=0.7\textwidth]{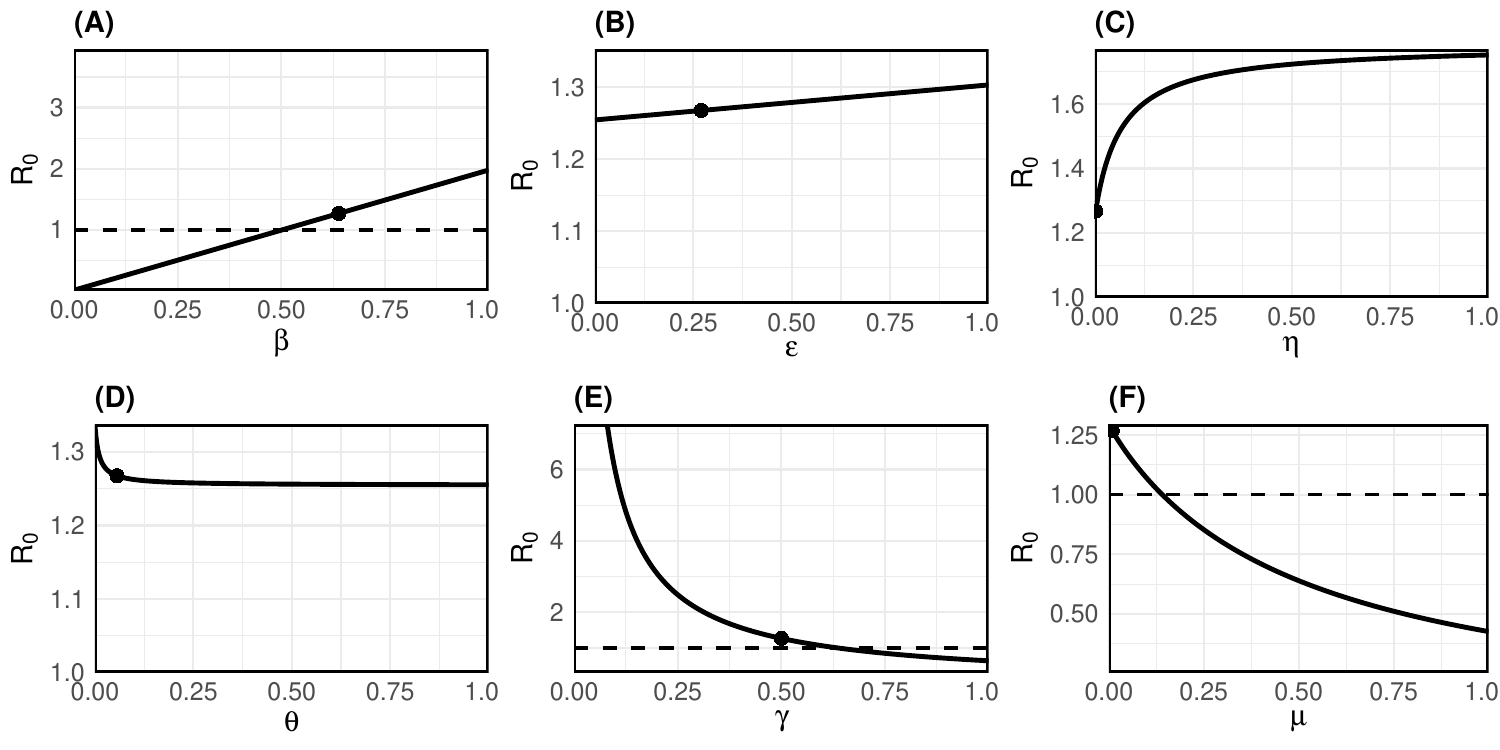}
\caption{Scenario analysis of the basic reproduction number $\mathcal{R}_0$ as a function of
selected parameters. The dashed line indicates the epidemic threshold
$\mathcal{R}_0 = 1$. The solid black disks indicate the baseline parameter values.Parameter values are taken from
\ref{table:params}}
\label{fig:scenario}
\end{figure}

To complement the sensitivity indices, we evaluated $\mathcal{R}_0$ as a
function of one parameter at a time, holding all others at their baseline values
(Table~\ref{table:params}). At baseline, $\mathcal{R}_0 \approx 1.267$, so the
disease-free equilibrium is unstable and the infection persists. Figure~\ref{fig:scenario} illustrates confirms scenario analysis of $\mathcal{R}_0$ : $\mathcal{R}_0$ increases linearly with $\beta$ (A) and $\varepsilon$
(B) and monotonically with the adoption rate $\eta$ (C), decreases with the
cessation rate $\theta$ (D) and, most strongly, with the recovery rate $\gamma$
(E), and decreases with natural mortality $\mu$ (F). $\mathcal{R}_0$ falls below
unity when $\beta < 0.50$, $\gamma > 0.64$, or $\mu > 0.14$. In contrast,
varying $\varepsilon$, $\eta$, or $\theta$ alone cannot bring $\mathcal{R}_0$
below unity, because direct transmission by itself gives
$\beta/(\gamma+\mu) \approx 1.25 > 1$; behavioral interventions therefore need
to be combined with measures that reduce direct transmission or shorten the
infectious period.

\subsubsection{Behavioral parameter thresholds}
\label{subsec:R0_behavior}

The condition $\mathcal{R}_0<1$ for disease elimination can be written as
\begin{equation}
\beta(\theta+\mu)+(\beta+\varepsilon)\eta <
(\eta+\theta+\mu)(\gamma+\mu),
\label{R0_inequality_1}
\end{equation}

which, solved for $\theta$, $\eta$, and $\varepsilon$ in turn, yields three critical values obtained from the same surface $\mathcal{R}_0(\theta,\eta,\varepsilon)=1$ and hence not independent substituting any one back into \eqref{R0_inequality_1} recovers the other two:
\begin{align}
\theta_c &=
\frac{\eta(\beta+\varepsilon-\gamma-\mu)+\mu(\beta-\gamma-\mu)}
{\gamma+\mu-\beta},
\label{theta_c}\\[1mm]
\eta_c &=
\frac{(\gamma+\mu-\beta)(\theta+\mu)}
{\beta+\varepsilon-\gamma-\mu},
\label{eta_c}\\[1mm]
\varepsilon_c &=
\frac{(\gamma+\mu-\beta)(\eta+\theta+\mu)}{\eta}.
\label{eps_c}
\end{align}
We focus on the epidemiologically relevant regime $\gamma+\mu>\beta$, where direct transmission alone cannot sustain the epidemic and behavioral interventions can meaningfully affect control. Here $\varepsilon_c>0$ always; $\eta_c>0$ additionally requires $\beta+\varepsilon>\gamma+\mu$; and $\theta_c>0$ additionally requires $\eta(\beta+\varepsilon-\gamma-\mu)+\mu(\beta-\gamma-\mu)>0$, i.e., sufficient indirect risk relative to adoption and mortality.

Each threshold marks a maximum tolerable (or, for $\theta_c$, minimum required) rate for elimination: cessation must exceed $\theta_c$, adoption must stay below $\eta_c$, and indirect transmission below $\varepsilon_c$, or the epidemic persists despite the other controls. As $\eta\to0^+$, $\varepsilon_c\to\infty$, since indirect risk is irrelevant once the reuse compartment is essentially unpopulated.

Since the fitted value $\beta = 0.639$ (Section~\ref{validation}) exceeds
$\gamma + \mu = 0.509$, direct transmission alone sustains the epidemic at
baseline, and none of the behavioral thresholds exists. To illustrate the
thresholds, we therefore take $\beta = 0.505$, which lies in the regime
$\gamma+\mu>\beta$ and represents a setting in which direct transmission has
been reduced, for example through treatment; all other parameters are as in
Table~\ref{table:params}. With these values, $\mathcal{R}_0 \approx 1.005$, so
the infection persists, and the critical values are $\theta_c \approx 0.100$,
$\eta_c \approx 0.00098$, and $\varepsilon_c \approx 0.162$.

\begin{figure}[H]
\centering
\includegraphics[width=0.7\textwidth]{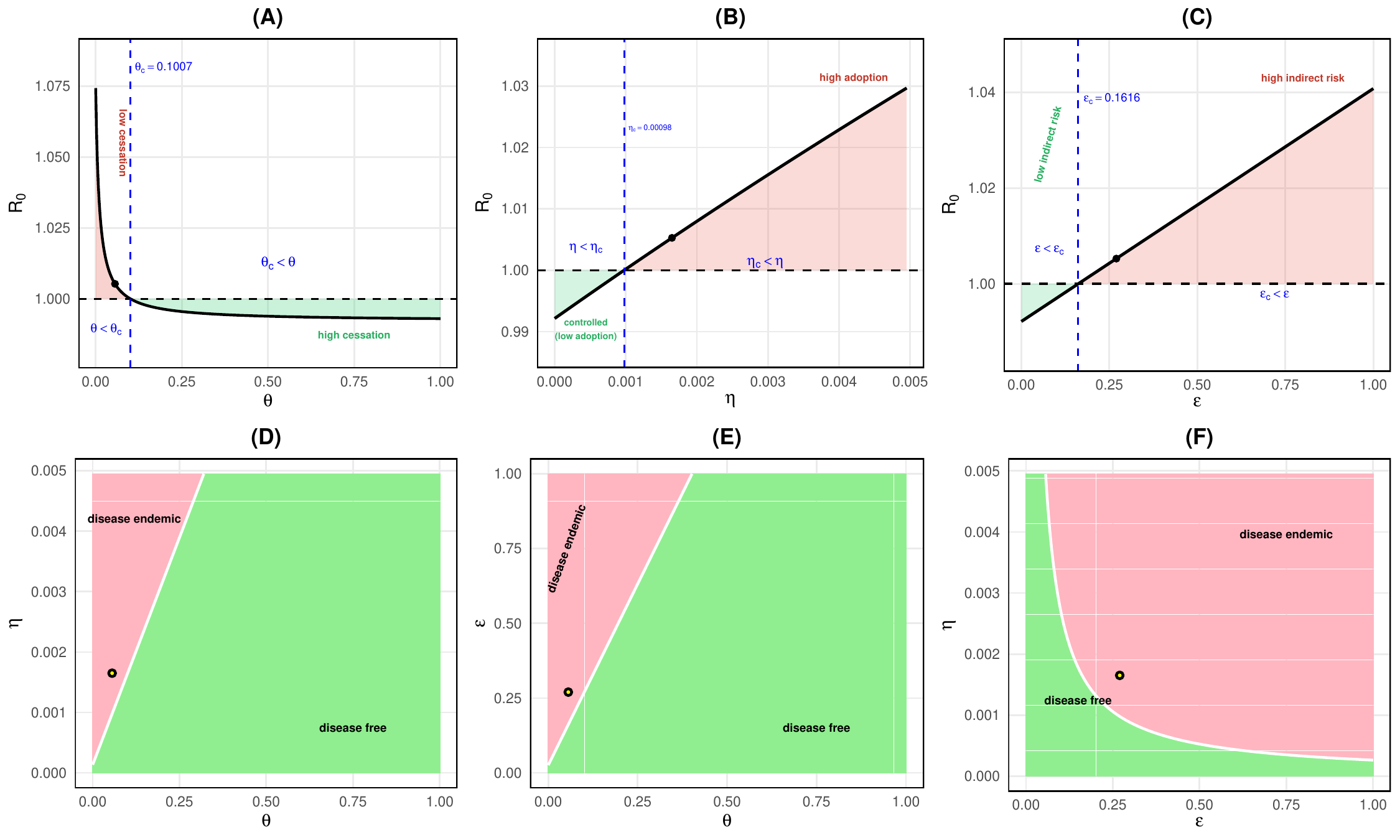}
\caption{Behavioral threshold dynamics of $\mathcal{R}_0$ in the SIRU model.
\textbf{(A)} $\mathcal{R}_0$ versus cessation rate $\theta$, with critical value $\theta_c$.
\textbf{(B)} $\mathcal{R}_0$ versus adoption rate $\eta$, with critical value $\eta_c$.
\textbf{(C)} $\mathcal{R}_0$ versus indirect transmission coefficient $\varepsilon$, with critical value $\varepsilon_c$.
\textbf{(D)} Parameter space $(\theta, \eta)$ with $\varepsilon$ fixed at its baseline value, divided by the $\mathcal{R}_0 = 1$ contour into disease-free and endemic regions.
\textbf{(E)} Parameter space $(\theta, \varepsilon)$ with $\eta$ fixed at its baseline value.
\textbf{(F)} Parameter space $(\varepsilon, \eta)$ with $\theta$ fixed at its baseline value.
In panels (A)--(C), the black disk marks the baseline value of $\mathcal{R}_0$; in panels (D)--(F), the black-outlined yellow disk marks the baseline parameter combination.
Parameter values taken from Table~\ref{table:params} except $\beta = 0.505$.}
\label{fig:R0_theta_eta}
\end{figure}

Figure~\ref{fig:R0_theta_eta}(A)--(C) show $\mathcal{R}_0$ crossing unity at
$\theta_c$, $\eta_c$, and $\varepsilon_c$, separating disease-free (green) from
endemic (red) regimes. Elimination requires increasing the cessation rate from
its baseline value of 0.056 to above $\theta_c$, roughly doubling it (A);
reducing the adoption rate from 0.00165 to below $\eta_c$, a reduction of about
40\% (B); or reducing the indirect transmission coefficient from 0.27 to below
$\varepsilon_c$, also a reduction of about 40\% (C). Panels (D)--(F) show the
corresponding two-parameter planes, each divided by the $\mathcal{R}_0=1$
contour. In each panel the baseline lies in the endemic region close to the
boundary, so that moderate changes in either behavioral parameter, or smaller
simultaneous changes in both, are sufficient to bring $\mathcal{R}_0$ below
unity.

These thresholds identify complementary intervention strategies: reducing adoption of syringe reuse ($\eta<\eta_c$) through prevention, increasing cessation ($\theta>\theta_c$) through harm-reduction and treatment interventions, and reducing the indirect transmission risk associated with syringe reuse ($\varepsilon<\varepsilon_c$), for instance through needle-exchange programs or improved syringe sterilization/disposal practices that lower the probability of using a contaminated syringe. These behavioral interventions are effective only once direct transmission has been brought close to its threshold value, which motivates combining them with measures, such as treatment, that reduce direct transmission.

\subsection{Local stability analysis of the DFE}

Using the DFE, \(E_0\) given in \eqref{E0}, its components satisfy
\[
N_0 = S_0 + U_0 = \frac{\Lambda}{\mu}.
\]

\begin{theorem}\label{thm:DFE-stability}
The disease-free equilibrium $E_0$ of the model is locally asymptotically stable if $\mathcal{R}_0 < 1$, and unstable if $\mathcal{R}_0 > 1$.
\end{theorem}

\begin{proof}
The local stability of $E_0$ is determined by the eigenvalues of the Jacobian matrix of the system evaluated at $E_0$. The Jacobian matrix is
\[
J(E_0)=
\begin{pmatrix}
-(\eta+\mu) & \theta & -\beta \dfrac{S_0}{N_0} & \alpha \\[2mm]
\eta & -(\theta+\mu) & -(\beta+\varepsilon)\dfrac{U_0}{N_0} & 0 \\[2mm]
0 & 0 & \beta + \varepsilon\dfrac{U_0}{N_0} - (\gamma+\mu) & 0 \\[2mm]
0 & 0 & \gamma & -(\alpha+\mu)
\end{pmatrix}.
\]

Since $J(E_0)$ is block-triangular, its eigenvalues are the diagonal entries:
\[
\lambda_1 = -(\eta+\mu), \quad \lambda_2 = -(\theta+\mu), \quad \lambda_3 = -(\alpha+\mu),
\]
and
\[
\lambda_4 = \beta + \varepsilon\frac{U_0}{N_0} - (\gamma+\mu).
\]

Using the expression for $U_0/N_0 = \eta/(\eta+\theta+\mu)$, we obtain
\[
\lambda_4 = \beta + \frac{\varepsilon \eta}{\eta+\theta+\mu} - (\gamma+\mu).
\]

The first three eigenvalues ($\lambda_1$, $\lambda_2$, and $\lambda_3$) are negative for all positive parameter values. The stability of $E_0$ thus depends on the sign of $\lambda_4$. Using the expression of $\mathcal{R}_0$  given in \eqref{R0}, we can write

\[
\lambda_4
= \beta + \frac{\varepsilon \eta}{\eta+\theta+\mu} - (\gamma+\mu)
= (\gamma+\mu)(\mathcal{R}_0 - 1).
\]

Therefore, $\lambda_4 < 0$ if and only if $\mathcal{R}_0 < 1$, and $\lambda_4 > 0$ if and only if $\mathcal{R}_0 > 1$. By the linearization principle, $E_0$ is locally asymptotically stable when $\mathcal{R}_0 < 1$ and unstable when $\mathcal{R}_0 > 1$.
\end{proof}

\section{Endemic equilibrium: Existence and stability}

When $\mathcal{R}_0>1$, we expect the disease to persist at an endemic level. The following theorem establishes the existence and uniqueness of such an equilibrium.

\begin{theorem}[Existence and uniqueness of endemic equilibrium]
\label{thm:endemic}
System~\eqref{SIRU_model} admits a unique endemic equilibrium $E^*=(S^*,U^*,I^*,R^*)$ with $I^*>0$ if and only if $\mathcal{R}_0>1$.
\end{theorem}

\begin{proof}
At equilibrium, the system reduces to a quadratic equation in the infection prevalence $x=I^*/N^*$:
\begin{equation}
f(x) = ax^2 + bx + c = 0,
\label{eq:endemic_quadratic}
\end{equation}
where the coefficient $a$ is strictly negative for all admissible parameters, while
\begin{equation}
c = \mu(\gamma+\mu)\left(1+\frac{\theta+\mu}{\eta}\right)(\mathcal{R}_0-1).
\label{eq:endemic_c}
\end{equation}

When $\mathcal{R}_0>1$, we have $c>0$. Since $a<0$, the downward-opening parabola $f(x)$ has $f(0)=c>0$, ensuring two real roots of opposite signs. Feasibility requires $x \in (0, x_{\max})$, where $x_{\max}=(\alpha+\mu)/(\alpha+\mu+\gamma)$ is the maximum biologically admissible prevalence. The unique positive root $x^*\in(0,x_{\max})$ yields the endemic equilibrium components:
\begin{equation}
I^* = x^* N^*, \quad
R^* = \frac{\gamma}{\alpha+\mu} I^*, \quad
U^* = \frac{(\gamma+\mu)N^*}{D(x^*)}, \quad
S^* = \frac{(\beta+\varepsilon)x^*+\theta+\mu}{\eta}U^*,
\label{eq:endemic_components}
\end{equation}
where $D(x) = \frac{\beta}{\eta}\big((\beta+\varepsilon)x+\theta+\mu\big) + (\beta+\varepsilon)$ and $N^*=\Lambda/\mu$.

The resulting equilibrium satisfies $I^*>0$ and yields $S^*>0$, $U^*>0$, and $R^*>0$ from \eqref{eq:endemic_components}, hence it is biologically feasible.

When $\mathcal{R}_0\leq1$, we have $c\leq0$, so $f(0)\leq0$. Combined with $a<0$, this implies no positive root exists in the feasible interval, and thus no endemic equilibrium. The complete derivation is provided in Appendix~\ref{app:endemic}.
\end{proof}

The local stability of the endemic equilibrium determines whether small perturbations decay or amplify over time, indicating whether the endemic state is dynamically robust.

\begin{theorem}[Local stability of endemic equilibrium]
\label{thm:local_endemic}
If $\mathcal{R}_0>1$, then the endemic equilibrium $E^*$ is locally asymptotically stable.
\end{theorem}

\begin{proof}
The Jacobian matrix $J(E^*)$ of system~\eqref{SIRU_model} evaluated at the endemic equilibrium has characteristic polynomial
\begin{equation}
\lambda^4 + a_1\lambda^3 + a_2\lambda^2 + a_3\lambda + a_4 = 0.
\label{eq:char_poly}
\end{equation}
Using the equilibrium relations, we verify that all coefficients are strictly positive:
\[
a_1>0, \quad a_2>0, \quad a_3>0, \quad a_4>0,
\]
and the Routh--Hurwitz criterion
\begin{equation}
a_1a_2a_3 > a_3^2 + a_1^2a_4
\label{eq:routh_hurwitz}
\end{equation}
is satisfied. These conditions guarantee that all eigenvalues have strictly negative real parts, ensuring local asymptotic stability of $E^*$. The detailed calculations are presented in Appendix~\ref{app:stability}.
\end{proof}

Biologically, Theorems~\ref{thm:endemic} and~\ref{thm:local_endemic} establish that when $\mathcal{R}_0>1$, the infection persists at a unique stable endemic level determined by the balance between transmission, recovery, behavioral transitions, and demographic turnover. Small deviations from this equilibrium naturally decay, indicating that the endemic state is structurally robust to minor perturbations in initial conditions.

\subsection{Bifurcation analysis of endemic equilibria}

In the last section, we established the existence and stability of endemic equilibria. We now investigate how the endemicity depends on key parameters. Particularly, we study the sensitivity and qualitative behavior of the endemic infection $I^*$ and syringe users population $U^*$ as we vary transmission parameters $(\beta, \epsilon)$ and behavioral parameters $(\eta, \theta)$. 

\begin{figure}[H]
 \centering
   \centering \includegraphics[width=\textwidth]
   {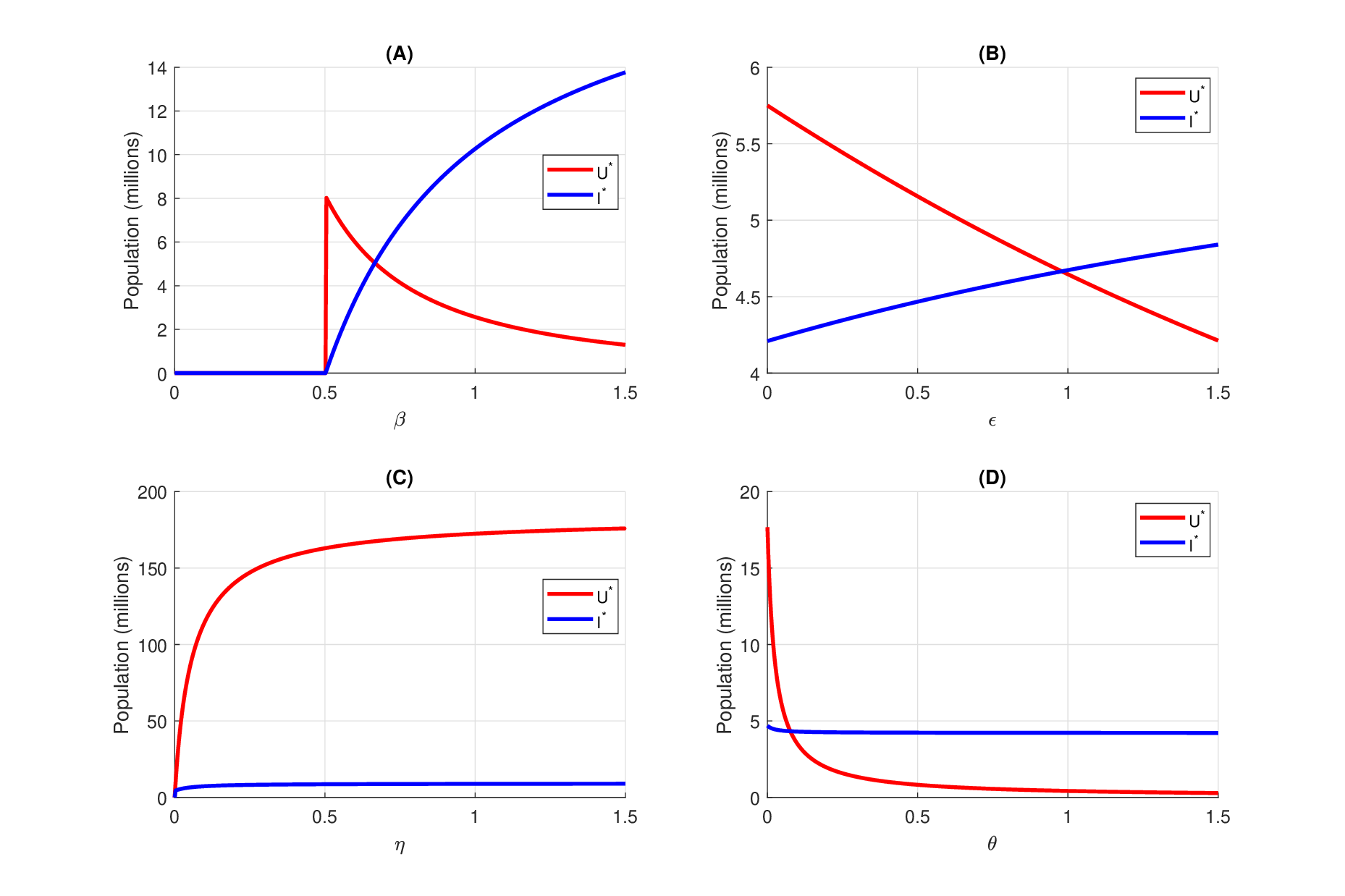}
   \caption{Endemic equilibrium populations $U^*$ and $I^*$ versus key parameters.
   } 
   \label{fig:bifurcation}
 \end{figure}

Figure~\ref{fig:bifurcation} shows how $I^*$ and $U^*$ vary with each parameter. Panel (A) exhibits a transcritical bifurcation at $\beta_c\approx0.5$ (where $\mathcal{R}_0=1$): the DFE is stable for $\beta<\beta_c$ and unstable for $\beta>\beta_c$, where a stable endemic state emerges. While $I^*$ transitions smoothly, $U^*$ jumps sharply at the bifurcation point -- a bottleneck effect from the small adoption rate $\eta$, which relaxes as reusers convert to infected individuals. Panels (B)--(D) show no such bifurcation: $I^*$ and $U^*$ increase smoothly and monotonically with the indirect transmission rate $\varepsilon$ (B, which acts as an additive contribution to transmission) and with the adoption rate $\eta$ (C), and decrease monotonically with the cessation rate $\theta$ (D). Because the baseline direct transmission $\beta=0.638$ alone exceeds $\beta_c$ and sustains the epidemic, the system remains endemic across the full range of $\eta$ and $\theta$, underscoring that behavioral interventions must be combined with reductions in direct transmission to achieve elimination.  

\section{Temporal variation of model dynamics}
\label{temporal}

To illustrate the dynamics of the model, the system~\eqref{SIRU_model} is solved numerically using the values of the baseline parameters in Table~\ref{table:params}. The recruitment rate is set to $\Lambda = \mu N(0)$, so that the total
population remains constant at its initial size. We take $N(0) = 311.6 \times 10^6$, the 2011 US population, with $U_0 = 3.7 \times 10^6$ syringe reusers \cite{bradley2023estimated}, $I_0 = 2.4 \times 10^6$ individuals living with HCV \cite{hofmeister2019estimating}, $R(0) = 0$, and $S_0 = N(0) - U_0 - I_0 = 305.5 \times 10^6$. The system is simulated over $[0, T]$ with $T = 20$ years. These simulations are illustrative scenarios of the model behavior rather than projections. We investigate the influence of the
key behavioral and transmission parameters by varying one parameter at a time over different values, while keeping all others fixed at their baseline values.

\begin{figure}[H]
\centering
\includegraphics[width=0.85\textwidth]{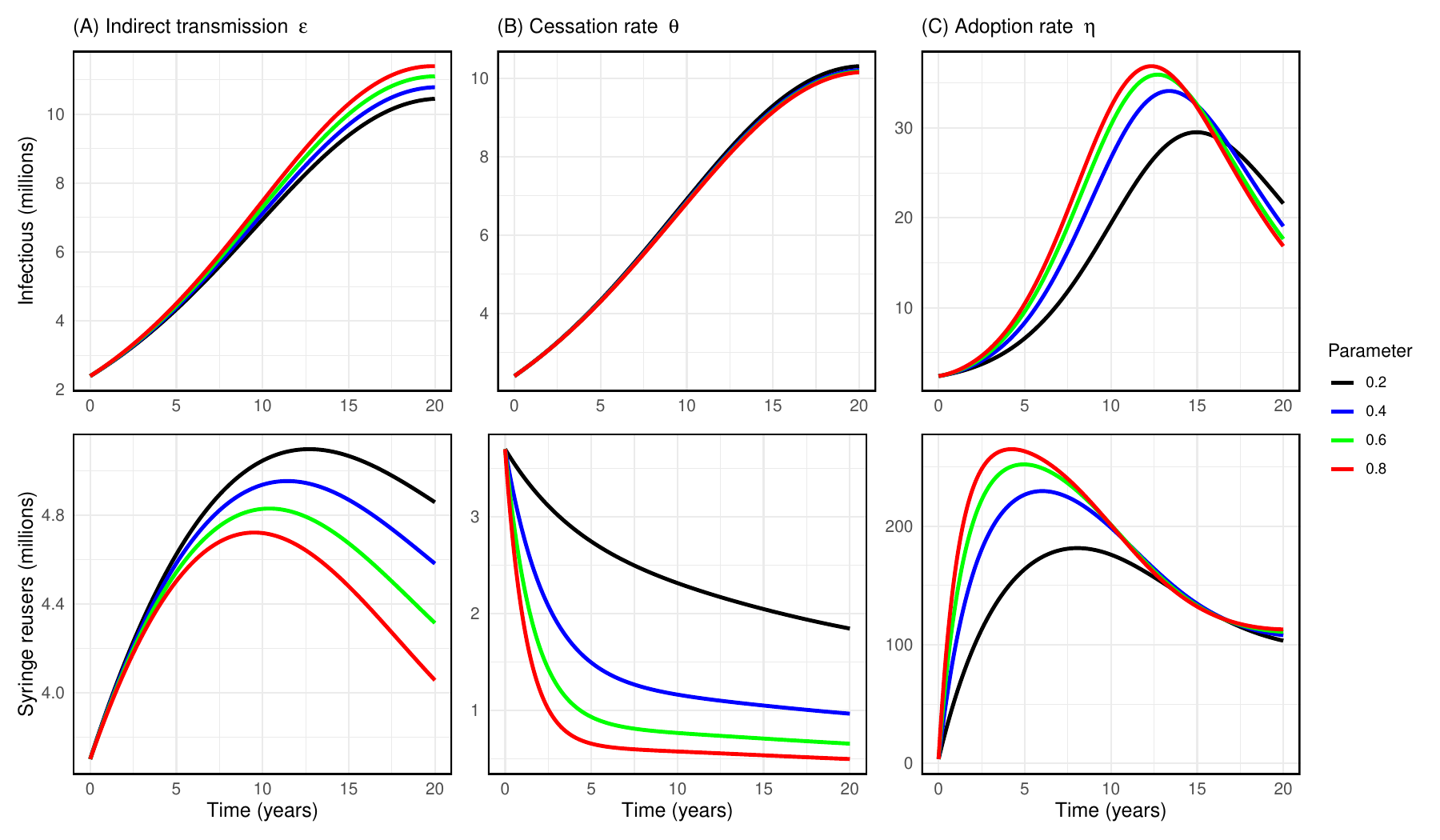}
\caption{Temporal dynamics of infectious individuals $I(t)$ (top row) and
syringe reusers $U(t)$ (bottom row) over 20 years for varying values of
(A) the indirect transmission coefficient $\varepsilon$, (B) the cessation rate
$\theta$, and (C) the adoption rate $\eta$. All other parameter values are taken
from Table~\ref{table:params}, with initial conditions
$(S_0, U_0, I_0, R(0)) = (305.5\text{ M}, 3.7\text{ M}, 2.4\text{ M}, 0)$.}
\label{fig:temporal_combined}
\end{figure}

Figure~\ref{fig:temporal_combined} shows the resulting dynamics of $I(t)$ (top row) and $U(t)$ (bottom row). Increasing the indirect transmission coefficient $\varepsilon$ (A) moderately increases infections, with $I(20)$ rising from about $10.4 \times 10^6$ to $11.4 \times 10^6$. The syringe reuser population
first grows, reaching a maximum of about $5 \times 10^6$, and then declines to between $4.9 \times 10^6$ and $4.1 \times 10^6$ after 20 years; a higher $\varepsilon$ lowers this maximum and brings it earlier, because it speeds the flow from $U$ to $I$ through contaminated syringes. This risk could be reduced through needle-syringe programs and access to sterile equipment. Increasing the
cessation rate $\theta$ (B) rapidly and substantially reduces the syringe reuser population, to between $1.9 \times 10^6$ and $0.5 \times 10^6$ after 20 years, but has only a small effect on total infections, underscoring that harm-reduction strategies that encourage cessation mainly shrink the high-risk
group. Increasing the adoption rate $\eta$ (C) has the strongest effect: the reuser population expands rapidly to a maximum of between $181 \times 10^6$ and
$265 \times 10^6$ within the first decade, and infections rise to a pronounced
peak of $30$--$37 \times 10^6$ that occurs earlier and is higher for larger
$\eta$. Both populations then decline as the susceptible pool is depleted.
Because these values of $\eta$ are well above the baseline adoption rate,
panel (C) represents high-adoption scenarios, and it shows that preventing entry
into syringe reuse is the most effective behavioral strategy, since it blocks
entry into the high-risk group before infection occurs.

\section{Optimal control and cost effectiveness analysis}
\label{sec:optimal_control}
We formulate an optimal control problem to identify time-dependent intervention strategies that minimize both HCV transmission and syringe reuse behavior in the controlled $SIR+U$ model. Two bounded control functions $u_1(t),u_2(t)\in[0,1]$ represent feasible public-health interventions implemented over a fixed intervention period $[0,T]$. The control $u_1(t)$ reduces the initiation of syringe reuse, while $u_2(t)$ enhances cessation of reuse behavior.

Specifically, we model these interventions by the modifications
\[
\eta \longrightarrow \eta(1-u_1(t)),\qquad 
\theta \longrightarrow \theta(1+u_2(t)),\qquad 0\le u_1(t),u_2(t)\le 1.
\]
 
The controlled system is given by
\begin{equation}
\label{eq:controlled_SIRU}
\begin{aligned}
\dfrac{dS}{dt} &= \Lambda - \beta \dfrac{I}{N} S - \eta(1-u_1) S + \alpha R + \theta(1+u_2) U - \mu S,\\
\dfrac{dU}{dt} &= \eta(1-u_1) S - \beta \dfrac{I}{N} U - \varepsilon \dfrac{I}{N} U - \theta(1+u_2) U - \mu U,\\
\dfrac{dI}{dt} &= \beta \dfrac{I}{N} S + \beta \dfrac{I}{N} U + \varepsilon \dfrac{I}{N} U - (\gamma+\mu) I,\\
\dfrac{dR}{dt} &= \gamma I - \alpha R - \mu R,
\end{aligned}
\end{equation}
where $N(t)=S(t)+U(t)+I(t)+R(t)$.

The objective functional is
\begin{equation}
\label{eq:objective}
J(u_1,u_2)=\int_0^T 
\Big[C_I I(t)+C_U U(t)+C_{u_1}u_1^2(t)+C_{u_2}u_2^2(t)\Big]\,dt.
\end{equation}
 
The optimal control problem consists of finding $(u_1^*,u_2^*)$ such that
\[
J(u_1^*,u_2^*)=\min_{(u_1,u_2)\in\mathcal U} J(u_1,u_2),
\qquad 
\mathcal U=\{(u_1,u_2): 0\le u_i(t)\le 1\}.
\]
 
To characterize the optimal controls, we apply Pontryagin’s Maximum Principle.

\begin{proposition}[Characterization of the optimal controls]
Let $(S^*,U^*,I^*,R^*)$ be an optimal state trajectory associated with an optimal control pair
$(u_1^*,u_2^*)$. Then there exist adjoint variables
$\lambda_S,\lambda_U,\lambda_I,\lambda_R$ satisfying
\[
\dot{\lambda}_k(t)=-\frac{\partial \mathcal H}{\partial k},\qquad 
\lambda_k(T)=0,\quad k\in\{S,U,I,R\},
\]
 
where $\mathcal H$ is the Hamiltonian defined in  Appendix \ref{Appendix_optimality}. Moreover, the optimal controls are given by
 
\begin{align}
u_1^*(t)
&=\min\!\left\{1,\max\!\left\{0,
\frac{\eta S^*(t)\big(\lambda_U(t)-\lambda_S(t)\big)}{2C_{u_1}}\right\}\right\},
\label{eq:optimal_u1_main}\\[2mm]
u_2^*(t)
&=\min\!\left\{1,\max\!\left\{0,
\frac{\theta U^*(t)\big(\lambda_U(t)-\lambda_S(t)\big)}{2C_{u_2}}\right\}\right\}.
\label{eq:optimal_u2_main}
\end{align}
\end{proposition}
The explicit form of the Hamiltonian and the adjoint system are provided in Appendix \ref{Appendix_optimality}.

\subsection{Numerical implementation}
The optimality system was solved numerically using the forward-backward sweep
method~\cite{lenhart2007optimal} over a 20-year intervention. The state
equations were integrated forward in time using the LSODA
method~\cite{petzold1983automatic}, while the adjoint equations were solved
backward in time with the specified terminal conditions. The controls were
iteratively updated from the characterization
formulas~\eqref{eq:optimal_u1_main} and~\eqref{eq:optimal_u2_main} and
projected onto the admissible interval $[0,1]$. Convergence was declared when
the maximum change in the control profiles between successive iterations was
less than $10^{-4}$. The simulations use the parameter values in
Table~\ref{table:params} and the same initial conditions as in
Section~\ref{temporal}.

\begin{figure}[H]
    \centering
    \includegraphics[width=0.8\textwidth]{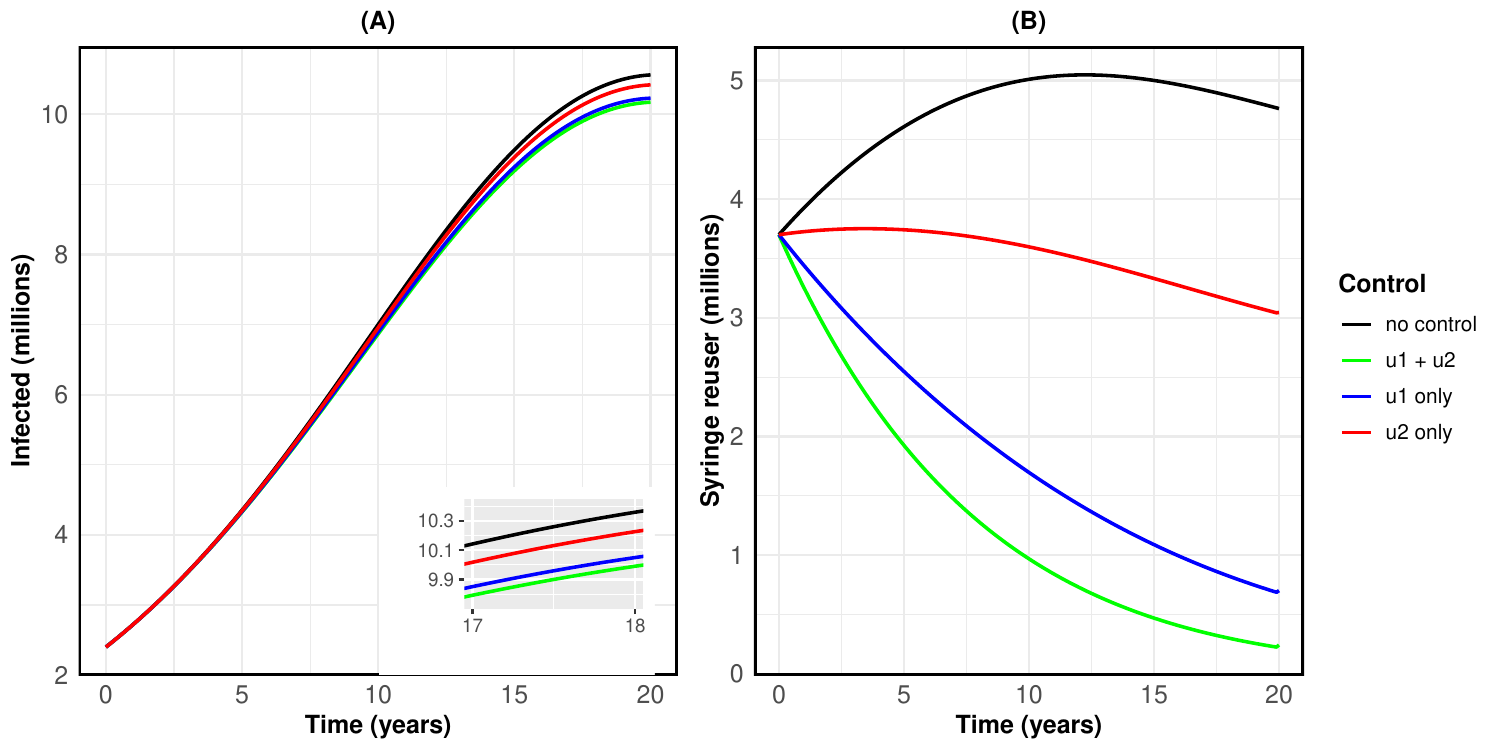}
    \caption{Temporal dynamics of (A) infected individuals and (B) syringe
    reusers under four control strategies over 20 years: no control,
    prevention of reuse initiation ($u_1$), cessation/harm reduction support
    ($u_2$), and combined controls ($u_1 + u_2$). Initial conditions are $S(0)=305.5$M, $U(0)=3.7$M, $I(0)=2.4$M, and $R(0)=0$. All
    model parameters are taken from Table~\ref{table:params}. Cost weights are
    $C_I = 1{,}000$, $C_U = 500$, $C_{u_1} = 75{,}000$, and
    $C_{u_2} = 75{,}000$.}
\label{figure:control}
\end{figure}

Figure~\ref{figure:control} compares infected individuals and syringe reusers
under the four control strategies. Prevention of reuse initiation
($u_1$) produces a rapid and substantial decline in the syringe reuser
population, reducing cumulative reuser person-years by 60.8\% over 20 years
relative to no control, while cessation alone ($u_2$) is more limited
(25.8\%), since it acts only after individuals have already entered the reuse
class. The combined strategy yields the largest reduction (72.9\%). The
corresponding reductions in cumulative infected person-years are much smaller,
1.9\% for $u_1$, 0.8\% for $u_2$, and 2.4\% for the combined strategy,
because syringe reusers form a small fraction of the population and most
transmission occurs through direct contact. In all scenarios, the optimal
controls remain close to their maximum level throughout most of the
intervention period and decrease only near the final time, reflecting the low
cost of the interventions relative to the burden of infection and syringe
reuse.

\subsection{Cost-effectiveness analysis}
Cost-effectiveness was evaluated by comparing each intervention strategy to the
no-control baseline. Health outcomes were measured by the 
person-years averted (TIA) and syringe-reuser person-years averted (TUA),
defined as reductions in the cumulative integrals of $I(t)$ and $U(t)$ over the
intervention period. Incremental cost-effectiveness ratios (ICERs) were
calculated as the additional cost per unit of health benefit:
{\small
\begin{align*}
\text{ICER}_{\text{TIA}} &= \frac{\text{Total Cost}_{\text{intervention}} - \text{Total Cost}_{\text{baseline}}}{\text{TIA}}, \\
\text{ICER}_{\text{TUA}} &= \frac{\text{Total Cost}_{\text{intervention}} - \text{Total Cost}_{\text{baseline}}}{\text{TUA}}.
\end{align*}
}

Table~\ref{table:CEA} summarizes the economic and epidemiological outcomes. All
three intervention strategies yield lower total cost than the no-control
baseline, producing negative ICER values and confirming that each is
cost-saving (dominant) relative to no control for the parameter values
considered. The prevention-only strategy $u_1$ achieves a large
reduction in syringe reuse and a smaller reduction in infections, with total
savings of $3.15\times10^{10}$, while the cessation-only strategy $u_2$ is
less effective, saving $1.34\times10^{10}$. The combined strategy provides the
greatest overall health benefit and the lowest total cost, with savings of
$3.78\times10^{10}$, highlighting the value of prioritizing prevention of
syringe reuse initiation while complementing it with cessation support. Because
these behavioral interventions act mainly on the reuser population, their
effect on overall HCV incidence remains modest, which reinforces the need to
combine them with measures that reduce direct transmission, such as
treatment.

\begin{table}[H]
\caption{Cost-effectiveness comparison of intervention strategies relative to
the no-control baseline over 20 years.
\label{table:CEA}}
\begin{tabular*}{\columnwidth}{@{\extracolsep\fill}lccccc@{\extracolsep\fill}}
\toprule
\textbf{Strategy} & \textbf{Total Cost} & \textbf{TIA} & \textbf{TUA} & \textbf{ICER (TIA)} & \textbf{ICER (TUA)} \\
\midrule
No control  & $1.85\times10^{11}$ & -- & -- & -- & -- \\
$u_1$       & $1.53\times10^{11}$ & $2.67\times10^{6}$ & $5.77\times10^{7}$ & $-1.18\times10^{4}$ & $-5.46\times10^{2}$ \\
$u_2$       & $1.71\times10^{11}$ & $1.14\times10^{6}$ & $2.45\times10^{7}$ & $-1.17\times10^{4}$ & $-5.47\times10^{2}$ \\
$u_1 + u_2$ & $1.47\times10^{11}$ & $3.27\times10^{6}$ & $6.91\times10^{7}$ & $-1.16\times10^{4}$ & $-5.47\times10^{2}$ \\
\bottomrule
\end{tabular*}
\vspace{0.2cm}
\footnotesize{TIA: infected person-years averted; TUA: syringe-reuser
person-years averted. Negative ICER values indicate cost-saving strategies.}
\end{table}

\subsection{Model fitted to incidence data}
\label{validation}

To assess the plausibility of the model against real-world data, we fitted it
to CDC estimates of annual new HCV infections in the United States over
2011--2021 \cite{cdc2018hepatitis, cdc2019hepatitis, cdc2020hepatitis,
cdc2021hepatitis}.

\begin{figure}[H]
    \centering
    \includegraphics[width=\linewidth]{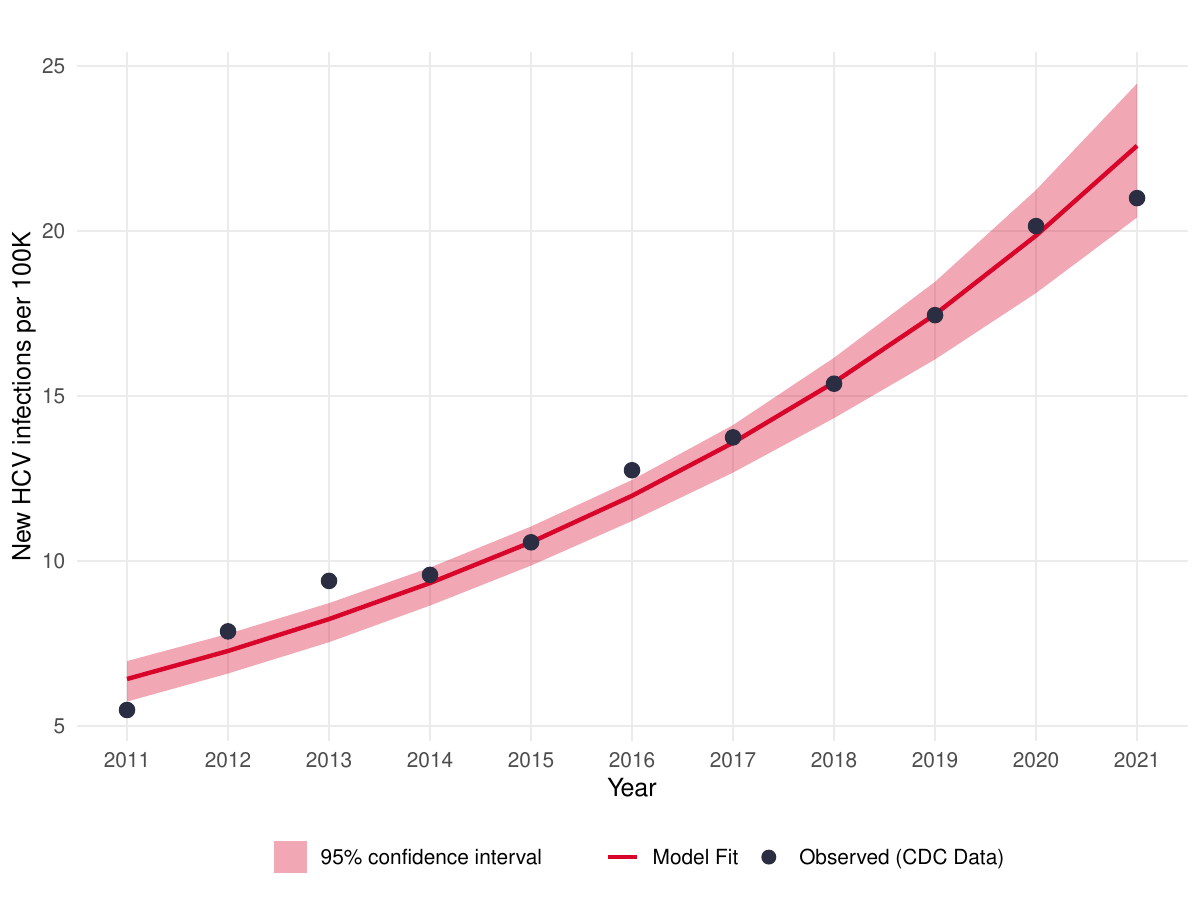}
    \caption{Model-predicted versus CDC-estimated annual new HCV infections per
    100,000 US population, 2011--2021. The shaded region shows the 95\%
    confidence interval of the fitted trajectory (parametric bootstrap,
    500 replicates).}
    \label{Fig:validation}
\end{figure}

With all other parameters as in Table~\ref{table:params}, we estimated the
direct transmission coefficient $\beta$ and the initial number of infectious
individuals $I_0$ by nonlinear least squares, minimizing the squared relative
differences between model-predicted and observed infections. Uncertainty was
quantified by parametric bootstrap (500 replicates). The estimated values are
$\beta = 0.639$ (95\% CI: 0.622--0.654) and $I_0 = 29{,}298$ (95\% CI:
25{,}407--33{,}021), giving $\mathcal{R}_0 = 1.268$ (95\% CI: 1.235--1.297).
As shown in Figure~\ref{Fig:validation}, the fitted model closely tracks the
observed data, and all observed values lie within the 95\% prediction interval.
Overall, these results support the model's ability to capture HCV transmission
dynamics in the United States.

\section{Discussion}

In this study, we developed a compartmental model of HCV transmission that explicitly incorporates syringe reuse as a behavioral process among PWID, distinguishing direct transmission from indirect, syringe-mediated infection rather than aggregating injecting risk into a single parameter. This sets our model apart from the majority of compartmental HCV frameworks reviewed by Martin et al.~\cite{Martin2018review}, which predominantly use susceptible infected(-susceptible) formulations without an explicit behavioral subgroup for high-risk injectors, and from the ecological framework of Miller-Dickson et al.~\cite{Miller2019}, which treats contaminated syringes as an environmental reservoir but not behavioral adoption or cessation as dynamic transitions.

The saturating bound on $\mathcal{R}_0$ we derive ($\mathcal{R}_0 \le (\beta+\varepsilon)/(\gamma+\mu)$; Section~\ref{subsec:R0}) implies that behavioral risk alone cannot drive unbounded epidemic growth transmission intensity and infectious duration must also be addressed. This is consistent with Scott et al.~\cite{ScottIceland2018}, who showed that antiviral scale-up alone reduced $\mathcal{R}_0$ from roughly 1.08 to below 0.6 in Iceland, and with the broader conclusion that neither harm reduction nor treatment alone is likely sufficient for elimination a combination of approaches targeting both transmission and infectious duration is required~\cite{Martin2018review}.

At the disease-free equilibrium, $\mathcal{R}_0$ reads as a weighted average of transmission across the susceptible and reuser subpopulations structurally similar to a classical SIR model, but with an explicit reuser compartment that grounds this effective transmission in concrete behavioral processes (adoption and cessation) rather than a single aggregated parameter. This shares the spirit of Tresna et al.~\cite{Tresna2022}, who stratified PWID by an educated/uneducated subgroup and found education-based interventions substantially reduced transmission; we instead track reuse behavior directly, yielding closed-form thresholds $\theta_c$ and $\eta_c$ that translate into concrete targets for harm-reduction programs.

Numerically, increasing $\beta$ accelerates and amplifies infection peaks, while reducing adoption ($\eta$) or increasing cessation ($\theta$) lowers the long-term burden by shrinking the high-risk subgroup; our bifurcation and temporal analyses further show convergence to a unique endemic equilibrium when $\mathcal{R}_0>1$ and to the DFE when $\mathcal{R}_0<1$, providing strong numerical (though not analytically proven global) support for these equilibria. The dominant sensitivity of $\mathcal{R}_0$ to the recovery rate $\gamma$ echoes the treatment-as-prevention literature: Blake and Smith~\cite{Blake2021} found improving treatment uptake had the largest single impact on HCV prevalence among PWID in New Hampshire, and Stocks et al.~\cite{stocks2020} similarly identified infectious duration as a dominant driver of reproduction-number estimates across European settings.

Our optimal control analysis shows that preventing reuse initiation is more effective than cessation alone, since it blocks entry into the high-risk group before exposure occurs; combining both yields the greatest overall reduction in reuse prevalence and infection burden. This hierarchy combined strategies outperforming single-component ones mirrors Marquez et al.~\cite{Marquez2020} in Tijuana, where combined antiviral scale-up and harm-reduction expansion was needed to meet WHO elimination targets, and Bellerose et al.~\cite{Bellerose2021}, whose review of network models found that pairing treatment with structural harm-reduction interventions maximized population-level benefit; the cost-saving nature of all three strategies (negative ICERs) further supports investing in behavioral prevention alongside biomedical treatment.

 Our model assumes homogeneous mixing and random syringe reuse within a mean-field framework, which does not capture network-level heterogeneity in real injection populations~\cite{Bellerose2021,Gicquelais2023}. Parameter values were drawn from multiple sources subject to uncertainty, so our numerical results should be read qualitatively rather than as precise quantitative predictions. In future work, we plan to extend the model by incorporating treatment dynamics, adherence, and relapse, as well as structured behavioral transitions such as rehabilitation programs. Such extensions would strengthen the connection between theoretical analysis, numerical investigation, and practical public health decision making.

\appendix

\section{Derivation of basic reproduction number}
\label{appendix_R0}

The basic reproduction number $\mathcal{R}_0$ is computed using the next-generation matrix approach of Diekmann et al.~\cite{Diekmann1990} and van den Driessche and Watmough~\cite{vandenDriessche2002}; since the model has a single infected compartment, the next-generation matrix reduces to a scalar.

The equation governing the infected class $I$ in system~\eqref{SIRU_model}
can be written as
\[
\frac{dI}{dt} = \mathcal{F}(S,U,I,R) - \mathcal{V}(S,U,I,R),
\]
where $\mathcal{F}$ is the rate of appearance of new infections,
\[
\mathcal{F}
= \beta \frac{I}{N} S
+ \beta \frac{I}{N} U
+ \varepsilon \frac{I}{N} U
= \frac{I}{N}\big[\beta S + (\beta+\varepsilon)U\big],
\]
and $\mathcal{V}$ collects all other transfers, $\mathcal{V} = (\gamma+\mu)I$. Following \cite{vandenDriessche2002}, the scalar matrices $F$ and $V$, evaluated at the disease-free equilibrium $E_0$, are
\[
F = \left[\frac{\partial \mathcal{F}}{\partial I}\right]_{E_0}
= \frac{1}{N^0}\big[\beta S^0 + (\beta+\varepsilon)U^0\big],
\qquad
V = \left[\frac{\partial \mathcal{V}}{\partial I}\right]_{E_0}
= \gamma+\mu,
\]
giving the basic reproduction number
\begin{equation}
\mathcal{R}_0 = \rho(FV^{-1})
= \frac{\beta S^0 + (\beta+\varepsilon)U^0}{N^0(\gamma+\mu)}.
\label{app:R0_general}
\end{equation}

Substituting the expressions for $S^0$, $U^0$, and $N^0$ into
\eqref{app:R0_general} and simplifying yields
\begin{equation*}
\mathcal{R}_0
=
\frac{\beta}{\gamma+\mu}
+
\frac{\varepsilon\,\eta}{(\theta+\eta+\mu)(\gamma+\mu)}.
\end{equation*}
\section{Endemic equilibrium derivation}
\label{app:endemic}

Let $E^*=(S^*,U^*,I^*,R^*)$ be an endemic equilibrium with $I^*>0$, satisfying
\begin{align}
0 &= \Lambda - \beta \tfrac{I^*}{N^*} S^* - \eta S^* + \alpha R^* + \theta U^* - \mu S^*, \label{eq:endemic_eq1} \\
0 &= \eta S^* - (\beta+\varepsilon) \tfrac{I^*}{N^*} U^* - \theta U^* - \mu U^*, \label{eq:endemic_eq2} \\
0 &= \beta \tfrac{I^*}{N^*} (S^*+U^*) + \varepsilon \tfrac{I^*}{N^*} U^* - (\gamma + \mu) I^*, \label{eq:endemic_eq3} \\
0 &= \gamma I^* - (\alpha+\mu) R^*. \label{eq:endemic_eq4}
\end{align}

From \eqref{eq:endemic_eq4}, $R^* = \frac{\gamma}{\alpha+\mu} I^*$. Set $x=I^*/N^*>0$. Then \eqref{eq:endemic_eq2} gives
\begin{equation}
S^* = \frac{(\beta+\varepsilon)x + \theta + \mu}{\eta} U^*,
\label{eq:endemic_Sstar}
\end{equation}
and dividing \eqref{eq:endemic_eq3} by $I^*$, substituting \eqref{eq:endemic_Sstar}, and solving for $U^*$ gives
\begin{equation}
U^* = \frac{(\gamma+\mu) N^*}{D(x)}, \qquad
D(x) = \frac{\beta}{\eta}\big((\beta+\varepsilon)x + \theta + \mu\big) + (\beta+\varepsilon).
\label{eq:endemic_Ustar}
\end{equation}

Since $S^*+U^*+I^*+R^*=N^*$, positivity of $S^*,U^*$ requires
\begin{equation}
x < \frac{\alpha+\mu}{\alpha+\mu+\gamma} =: x_{\max}.
\label{eq:endemic_xmax}
\end{equation}

Substituting $\Lambda=\mu N^*$, $I^*=xN^*$, and \eqref{eq:endemic_Sstar}--\eqref{eq:endemic_Ustar} into \eqref{eq:endemic_eq1}, dividing by $N^*$ and multiplying by $D(x)$, yields the quadratic
\begin{equation}
f(x) = ax^2+bx+c=0,
\label{eq:endemic_quadratic_full}
\end{equation}
with, writing $K_1=\frac{\beta(\beta+\varepsilon)}{\eta}$, $K_2=\frac{\beta(\theta+\mu)}{\eta}+(\beta+\varepsilon)$,
\begin{align}
a &= K_1\Big(\tfrac{\alpha\gamma}{\alpha+\mu}-(\gamma+\mu)\Big), \label{eq:endemic_coeff_a} \\
b &= \mu K_1 + \tfrac{\alpha\gamma}{\alpha+\mu}K_2 - \tfrac{\gamma+\mu}{\eta}\big[\beta(\theta+\mu)+(\eta+\mu)(\beta+\varepsilon)\big], \label{eq:endemic_coeff_b} \\
c &= \mu K_2 + (\gamma+\mu)\Big[\theta - \tfrac{(\eta+\mu)(\theta+\mu)}{\eta}\Big]. \label{eq:endemic_coeff_c}
\end{align}
Factoring $\frac{\mu(\eta+\theta+\mu)}{\eta}$ from \eqref{eq:endemic_coeff_c} and using
$\mathcal{R}_0 = \frac{\beta}{\gamma+\mu}+\frac{\varepsilon\eta}{(\eta+\theta+\mu)(\gamma+\mu)}$ gives
\begin{equation}
c = \mu(\gamma+\mu)\Big(1+\tfrac{\theta+\mu}{\eta}\Big)(\mathcal{R}_0-1).
\label{eq:endemic_c_R0}
\end{equation}

Since $\frac{\alpha\gamma}{\alpha+\mu}<\gamma+\mu$, \eqref{eq:endemic_coeff_a} gives $a<0$, so $f$ opens downward. If $\mathcal{R}_0>1$, then $c>0=f(0)>0$; with $a<0$ and $f(x)\to-\infty$, the IVT gives a positive root, unique since $c/a<0$, feasible as $x^*\in(0,x_{\max})$. If $\mathcal{R}_0\le1$, then $c\le0$, so $f(0)\le0$ and, with $a<0$, $f$ has no positive root in $(0,x_{\max})$: no endemic equilibrium exists. This completes the proof of Theorem~\ref{thm:endemic}. \qed

\section{Local stability of the endemic equilibrium}
\label{app:stability}

The Jacobian of system~\eqref{SIRU_model} at $E^*=(S^*,U^*,I^*,R^*)$, well-defined since $N^*=\Lambda/\mu$ and all components of $E^*$ are strictly positive when $\mathcal{R}_0>1$, is
\[
J(E^*) =
\begin{pmatrix}
-\beta \dfrac{I^*}{N^*} - \eta - \mu &
\theta &
-\beta \dfrac{S^*}{N^*} &
\alpha \\[8pt]
\eta &
-(\beta+\varepsilon)\dfrac{I^*}{N^*} - \theta - \mu &
-(\beta+\varepsilon)\dfrac{U^*}{N^*} &
0 \\[8pt]
\beta \dfrac{I^*}{N^*} &
(\beta+\varepsilon)\dfrac{I^*}{N^*} &
\beta \dfrac{S^*+U^*}{N^*} - (\gamma+\mu) &
0 \\[8pt]
0 & 0 & \gamma & -(\alpha+\mu)
\end{pmatrix}.
\]
Its characteristic equation is
\begin{equation}
\lambda^4 + a_1 \lambda^3 + a_2 \lambda^2 + a_3 \lambda + a_4 = 0,
\label{eq:stability_char_poly}
\end{equation}
with $a_1=-\operatorname{tr}(J(E^*))$. Using $\beta\frac{S^*+U^*}{N^*}=\gamma+\mu-\varepsilon\frac{U^*}{N^*}$ from \eqref{eq:endemic_eq3},
\[
\operatorname{tr}(J(E^*))= -(2\beta+\varepsilon)\frac{I^*}{N^*} -\varepsilon\frac{U^*}{N^*} -(\eta+\theta+\alpha+3\mu) < 0,
\]
so $a_1>0$. Writing the entries of $J(E^*)$ as $j_{11}=-\beta\frac{I^*}{N^*}-\eta-\mu$, $j_{22}=-(\beta+\varepsilon)\frac{I^*}{N^*}-\theta-\mu$, $j_{33}=-\varepsilon\frac{U^*}{N^*}$, $j_{44}=-(\alpha+\mu)$ (all negative), with off-diagonal entries $j_{12}=\theta$, $j_{13}=-\beta\frac{S^*}{N^*}$, $j_{14}=\alpha$, $j_{21}=\eta$, $j_{23}=-(\beta+\varepsilon)\frac{U^*}{N^*}$, $j_{31}=\beta\frac{I^*}{N^*}$, $j_{32}=(\beta+\varepsilon)\frac{I^*}{N^*}$, $j_{43}=\gamma$, each $2\times2$ principal minor
\begin{equation*}
\begin{aligned}
M_{12} &= j_{11}j_{22}-j_{12}j_{21}, &
M_{13} &= j_{11}j_{33}-j_{13}j_{31}, &
M_{14} &= j_{11}j_{44},\\
M_{23} &= j_{22}j_{33}-j_{23}j_{32}, &
M_{24} &= j_{22}j_{44}, &
M_{34} &= j_{33}j_{44},
\end{aligned}
\end{equation*}
is positive, so $a_2=\sum_{i<j}M_{ij}>0$. Similarly, all $3\times3$ principal minors are negative by the sign structure of $J(E^*)$, giving $a_3=-\sum(\text{3 minors})>0$, and $a_4=\det(J(E^*))>0$.

For the Routh--Hurwitz condition
\begin{equation}
a_1 a_2 a_3 > a_3^2 + a_1^2 a_4,
\label{eq:stability_routh_hurwitz}
\end{equation}
note $J(E^*)$ is block-triangular,
\[
J(E^*) =
\begin{pmatrix}
\mathcal{J}_{3\times 3} & * \\
0 & -(\alpha+\mu)
\end{pmatrix},
\]
so $\det(\lambda I - J(E^*)) = (\lambda + \alpha + \mu)\det(\lambda I - \mathcal{J}_{3\times 3})$, and stability reduces to the $(S,U,I)$-subsystem. Since $\Delta_3=a_1a_2a_3-a_3^2-a_1^2a_4$ is continuous in the parameters, positive near the transcritical bifurcation at $\mathcal{R}_0=1$ (where $E^*$ emerges from the disease-free equilibrium), and $a_1,a_2,a_3,a_4>0$ throughout $\mathcal{R}_0>1$, it follows that $\Delta_3>0$ for all $\mathcal{R}_0>1$.

Hence all eigenvalues of $J(E^*)$ have negative real parts for $\mathcal{R}_0>1$, so $E^*$ is locally asymptotically stable, completing the proof of Theorem~\ref{thm:local_endemic}. \qed

\section{Derivation of the optimality system}
\label{Appendix_optimality}

The Hamiltonian for the optimal control problem of Section~\ref{sec:optimal_control} is
\begin{equation}
\label{eq:hamiltonian}
\begin{aligned}
\mathcal{H} &= C_I I + C_U U + C_{u_1} u_1^2 + C_{u_2} u_2^2  \\
&\quad + \lambda_S \Big[\Lambda - \beta \frac{I}{N} S - \eta(1-u_1) S + \alpha R + \theta(1+u_2) U - \mu S \Big] \\
&\quad + \lambda_U \Big[\eta(1-u_1) S - \beta \frac{I}{N} U - \varepsilon \frac{I}{N} U - \theta(1+u_2) U - \mu U \Big] \\
&\quad + \lambda_I \Big[\beta \frac{I}{N} S + \beta \frac{I}{N} U + \varepsilon \frac{I}{N} U - (\gamma+\mu) I \Big] \\
&\quad + \lambda_R \Big[\gamma I - \alpha R - \mu R \Big].
\end{aligned}
\end{equation}
By Pontryagin's Maximum Principle, the adjoint variables satisfy $\dot{\lambda}_k=-\partial \mathcal H/\partial k$, $\lambda_k(T)=0$, $k\in\{S,U,I,R\}$, giving
\begin{align}
\frac{d\lambda_S}{dt} &=
\lambda_S\!\left[ \beta \frac{I(N-S)}{N^2} + \eta(1-u_1) + \mu \right]
- \lambda_U\!\left[ \eta(1-u_1) + (\beta+\varepsilon)\frac{IU}{N^2} \right] \notag\\
&\quad
- \lambda_I\!\left[ \beta \frac{I(N-S)}{N^2}
- (\beta+\varepsilon)\frac{IU}{N^2} \right], \label{eq:adjS}\\[2mm]
\frac{d\lambda_U}{dt} &=
-C_U
- \lambda_S\!\left[ \beta \frac{IS}{N^2} + \theta(1+u_2) \right]
+ \lambda_U\!\left[ (\beta+\varepsilon)\frac{I(N-U)}{N^2}
+ \theta(1+u_2) + \mu \right] \notag\\
&\quad
+ \lambda_I\!\left[ \beta \frac{IS}{N^2}
- (\beta+\varepsilon)\frac{I(N-U)}{N^2} \right], \label{eq:adjU}\\[2mm]
\frac{d\lambda_I}{dt} &=
-C_I
+ \lambda_S\!\left[ \beta \frac{S(N-I)}{N^2} \right]
+ \lambda_U\!\left[ (\beta+\varepsilon)\frac{U(N-I)}{N^2} \right] \notag\\
&\quad
- \lambda_I\!\left[ \beta \frac{S(N-I)}{N^2}
+ (\beta+\varepsilon)\frac{U(N-I)}{N^2}
- (\gamma+\mu) \right]
- \lambda_R\gamma, \label{eq:adjI}\\[2mm]
\frac{d\lambda_R}{dt} &=
-\lambda_S\!\left[ \beta \frac{IS}{N^2} + \alpha \right]
- \lambda_U\!\left[ (\beta+\varepsilon)\frac{IU}{N^2} \right]
+ \lambda_I\!\left[ \beta \frac{IS}{N^2}
+ (\beta+\varepsilon)\frac{IU}{N^2} \right]
+ \lambda_R(\alpha+\mu). \label{eq:adjR}
\end{align}

\bibliographystyle{IEEEtran}
\bibliography{references}

@article{hayashi2018declining,
  title={Declining mortality rates in {HIV}-infected people who inject drugs during a seek-and-treat initiative in {Vancouver, Canada}, 1996--2014: a prospective cohort study},
  author={Hayashi, Kanna and Dong, Huiru and Kerr, Thomas and Dobrer, Sabina and Guillemi, Silvia and Barrios, Rolando and Montaner, Julio SG and Wood, Evan and Milloy, M-J},
  journal={The Journal of Infectious Diseases},
  volume={217},
  number={1},
  pages={64--68},
  year={2018},
  publisher={Oxford University Press}
}

@article{lansky2014,
  title={Estimating the number of persons who inject drugs in the {United States} by meta-analysis to calculate national rates of {HIV} and hepatitis {C} virus infections},
  author={Lansky, A. and Finlayson, T. and Johnson, C. and others},
  journal={PLoS One},
  volume={9},
  number={5},
  pages={e97596},
  year={2014}
}

@article{stark1997,
  title={History of syringe sharing in prison and risk of {Hepatitis B} virus, {Hepatitis C} virus, and human immunodeficiency virus infection among injecting drug users in {Berlin}},
  author={Stark, K. and Bienzle, U. and Vonk, R. and Guggenmoos-Holzmann, I.},
  journal={International Journal of Epidemiology},
  volume={26},
  number={6},
  pages={1359--1366},
  year={1997}
}

@article{elaiw2024global,
  title={Global analysis of {HCV/HBV} codynamics model with antibody immunity},
  author={Elaiw, Ahmed M. and Al Agha, Afnan D. and Alsaadi, Ghadeer and Hobiny, Atef D.},
  journal={The European Physical Journal Plus},
  volume={139},
  number={9},
  pages={850},
  year={2024},
  publisher={Springer}
}

@article{binka2015,
  title={Survival of {Hepatitis C} virus in syringes is dependent on the design of the syringe-needle and dead space volume},
  author={Binka, M. and Paintsil, E. and Patel, A. and Lindenbach, B. D. and Heimer, R.},
  journal={PLoS One},
  volume={10},
  number={11},
  year={2015}
}

@article{scheibe2019,
  title={Understanding {Hepatitis B}, {Hepatitis C} and {HIV} among people who inject drugs in {South Africa}: findings from a three-city cross-sectional survey},
  author={Scheibe, A. and Young, K. and Moses, L.},
  journal={Harm Reduction Journal},
  volume={16},
  number={28},
  year={2019}
}

@article{thorpe2002risk,
  title={Risk of hepatitis {C} virus infection among young adult injection drug users who share injection equipment},
  author={Thorpe, Lorna E. and Ouellet, Lawrence J. and Hershow, Ronald and Bailey, Susan L. and Williams, Ian T. and Williamson, John and Monterroso, Edgar R. and Garfein, Richard S.},
  journal={American Journal of Epidemiology},
  volume={155},
  number={7},
  pages={645--653},
  year={2002}
}

@article{paquette2018injection,
  title={Injection drug use, {HIV/HCV}, and related services in nonurban areas of the {United States}: a systematic review},
  author={Paquette, Catherine E. and Pollini, Robin A.},
  journal={Drug and Alcohol Dependence},
  volume={188},
  pages={239--250},
  year={2018}
}

@article{martin2018review,
  title={Mathematical modeling of hepatitis {C} virus ({HCV}) prevention among people who inject drugs: a review of the literature and insights for elimination strategies},
  author={Martin, Natasha K. and Vickerman, Peter and Grebely, Jason and Hickman, Matthew and Hutchinson, Sharon J. and Dore, Gregory J.},
  journal={Journal of Theoretical Biology},
  volume={481},
  pages={194--206},
  year={2019}
}

@article{scotticeland2018,
  title={Modelling the elimination of hepatitis {C} as a public health threat in {Iceland}: a goal attainable by 2020},
  author={Scott, Nick and Olafsson, Sigurdur and Gottfresson, Magnus and Tyrfingsson, Thorarinn and Runarsdottir, Valgerdur and Hansdottir, Ingvi and Hernandez, Urdur B. and Sigmundsdottir, Gudrun and Hellard, Margaret},
  journal={Journal of Hepatology},
  volume={68},
  number={5},
  pages={932--939},
  year={2018},
  publisher={Elsevier}
}

@article{marquez2020,
  title={Is hepatitis {C} virus ({HCV}) elimination achievable among people who inject drugs in {Tijuana, Mexico}? {A} modeling analysis},
  author={Marquez, Lara K. and Cepeda, Javier A. and Borquez, Annick and Strathdee, Steffanie A. and Gonzalez-Z{\'u}{\~n}iga, Patricia E. and Fleiz, Clara and Rafful, Claudia and Garfein, Richard S. and Kiene, Susan M. and Brodine, Stephanie and Martin, Natasha K.},
  journal={International Journal of Drug Policy},
  volume={80},
  pages={102758},
  year={2020},
  publisher={Elsevier}
}

@article{bellerose2021,
  title={A review of network simulation models of hepatitis {C} virus and {HIV} among people who inject drugs},
  author={Bellerose, Meghan and Zhu, Liyang and Hagan, Liesl M. and Thompson, William W. and Randall, Lydia M. and Malyuta, Yelena and Salomon, Joshua A. and Linas, Benjamin P.},
  journal={International Journal of Drug Policy},
  volume={88},
  pages={102580},
  year={2021},
  publisher={Elsevier}
}

@article{gicquelais2023,
  title={Injection partnership characteristics and {HCV} status associations with syringe and equipment sharing among people who inject drugs},
  author={Gicquelais, Rachel E. and Mackesy-Amiti, Mary Ellen and Boodram, Basmattee and Page, Kimberly and Latkin, Carl},
  journal={BMC Public Health},
  volume={23},
  pages={1191},
  year={2023},
  publisher={Springer}
}

@article{kermack1927,
  title={A contribution to the mathematical theory of epidemics},
  author={Kermack, W. O. and McKendrick, A. G.},
  journal={Proceedings of the Royal Society of London. Series A},
  volume={115},
  number={772},
  pages={700--721},
  year={1927},
  publisher={The Royal Society}
}

@book{martcheva2015,
  title={An Introduction to Mathematical Epidemiology},
  author={Martcheva, M.},
  year={2015},
  publisher={Springer},
  address={New York}
}

@article{sacks2012behavioural,
  title={Behavioural interventions for preventing hepatitis {C} infection in people who inject drugs: a global systematic review},
  author={Sacks-Davis, Rachel and Horyniak, Danielle and Grebely, Jason and Hellard, Margaret},
  journal={International Journal of Drug Policy},
  volume={23},
  number={3},
  pages={176--184},
  year={2012},
  publisher={Elsevier}
}

@article{hagan2013best,
  title={Best strategies for global {HCV} eradication},
  author={Hagan, Liesl M. and Schinazi, Raymond F.},
  journal={Liver International},
  volume={33},
  pages={68--79},
  year={2013},
  publisher={Wiley Online Library}
}

@article{kaplan1992model,
  title={A model-based estimate of {HIV} infectivity via needle sharing},
  author={Kaplan, Edward H. and Heimer, Robert},
  journal={Journal of Acquired Immune Deficiency Syndromes},
  volume={5},
  number={11},
  pages={1116--1118},
  year={1992}
}

@article{homer1991model,
  title={A model of {HIV} transmission through needle sharing},
  author={Homer, Jack B. and St. Clair, Christian L.},
  journal={Interfaces},
  volume={21},
  number={3},
  pages={26--49},
  year={1991},
  publisher={INFORMS}
}

@article{murray2003impact,
  title={The impact of behavioural changes on the prevalence of human immunodeficiency virus and hepatitis {C} among injecting drug users},
  author={Murray, John M. and Law, Matthew G. and Gao, Zhanhai and Kaldor, John M.},
  journal={International Journal of Epidemiology},
  volume={32},
  number={5},
  pages={708--714},
  year={2003},
  publisher={Oxford University Press}
}

@article{hutchinson2006modelling,
  title={Modelling the spread of hepatitis {C} virus infection among injecting drug users in {Glasgow}: implications for prevention},
  author={Hutchinson, Sharon J. and Bird, Sheila M. and Taylor, Avril and Goldberg, David J.},
  journal={International Journal of Drug Policy},
  volume={17},
  number={3},
  pages={211--221},
  year={2006},
  publisher={Elsevier}
}

@article{vickerman2006model,
  title={Model projections on the required coverage of syringe distribution to prevent {HIV} epidemics among injecting drug users},
  author={Vickerman, Peter and Hickman, Matthew and Rhodes, Tim and Watts, Charlotte},
  journal={Journal of Acquired Immune Deficiency Syndromes},
  volume={42},
  number={3},
  pages={355--361},
  year={2006},
  publisher={LWW}
}

@article{vickerman2007modelling,
  title={Modelling the impact on hepatitis {C} transmission of reducing syringe sharing: {London} case study},
  author={Vickerman, Peter and Hickman, Matthew and Judd, Alex},
  journal={International Journal of Epidemiology},
  volume={36},
  number={2},
  pages={396--405},
  year={2007},
  publisher={Oxford University Press}
}

@article{kwon2009impact,
  title={The impact of needle and syringe programs on {HIV} and {HCV} transmissions in injecting drug users in {Australia}: {A} model-based analysis},
  author={Kwon, Jisoo A. and Iversen, Jenny and Maher, Lisa and Law, Matthew G. and Wilson, David P.},
  journal={Journal of Acquired Immune Deficiency Syndromes},
  volume={51},
  number={4},
  pages={462--469},
  year={2009},
  publisher={LWW}
}

@article{martin2011mathematical,
  title={Mathematical modelling of hepatitis {C} treatment for injecting drug users},
  author={Martin, Natasha K. and Vickerman, Peter and Hickman, Matthew},
  journal={Journal of Theoretical Biology},
  volume={274},
  number={1},
  pages={58--66},
  year={2011},
  publisher={Elsevier}
}

@article{blake2021,
  title={Modeling hepatitis {C} elimination among people who inject drugs in {New Hampshire}},
  author={Blake, Andrew and Smith, James E.},
  journal={JAMA Network Open},
  volume={4},
  number={8},
  pages={e2119092},
  year={2021},
  publisher={American Medical Association}
}

@article{elbasha2013model,
  title={Model for hepatitis {C} virus transmissions},
  author={Elbasha, Elamin H.},
  journal={Mathematical Biosciences and Engineering},
  volume={10},
  number={4},
  pages={1045--1065},
  year={2013},
  publisher={American Institute of Mathematical Sciences}
}

@article{kretzschmar2004modelling,
  title={Modelling the impact of {HIV} and hepatitis {C} virus prevention among injecting drug users},
  author={Kretzschmar, Mirjam and Wiessing, Lucas},
  journal={Drug and Alcohol Dependence},
  volume={73},
  number={2},
  pages={159--171},
  year={2004},
  publisher={Elsevier}
}

@article{greenhalgh1997mathematical,
  title={Mathematical modelling of the spread of {HIV/AIDS} amongst injecting drug users},
  author={Greenhalgh, David and Hay, Gordon},
  journal={Mathematical Medicine and Biology: A Journal of the IMA},
  volume={14},
  number={1},
  pages={11--38},
  year={1997},
  publisher={OUP}
}

@article{white2007,
  title={Heroin epidemics, treatment, and {ODE} modelling},
  author={White, E. and Comiskey, C.},
  journal={Mathematical Biosciences},
  volume={208},
  number={1},
  pages={312--324},
  year={2007},
  publisher={Elsevier}
}

@article{scott2016,
  title={Modeling {Hepatitis C} virus transmission among people who inject drugs: assumptions, limitations and future challenges},
  author={Scott, N. and Hellard, M. and McBryde, E. S.},
  journal={Virulence},
  volume={7},
  number={2},
  pages={201--208},
  year={2016},
  publisher={Taylor \& Francis}
}

@article{miller2019,
  title={Hepatitis {C} virus modelled as an indirectly transmitted infection highlights the centrality of injection drug equipment in disease dynamics},
  author={Miller-Dickson, M. D. and Meszaros, V. A. and Almagro-Moreno, S. and Ogbunugafor, C. B.},
  journal={Journal of The Royal Society Interface},
  volume={16},
  number={158},
  pages={20190334},
  year={2019},
  publisher={The Royal Society}
}

@article{tresna2022,
  title={Mathematical model of {HCV} transmission with treatment and educational effort},
  author={Tresna, S. T. and Anggriani, N. and Supriatna, A. K.},
  journal={Communications in Mathematical Biology and Neuroscience},
  volume={2022},
  pages={Article ID 46},
  year={2022}
}

@article{bradley2023estimated,
  title={Estimated number of people who inject drugs in the {United States}},
  author={Bradley, Heather and Hall, Eric W. and Asher, Alice and Furukawa, Nathan W. and Jones, Christopher M. and Shealey, Jalissa and Buchacz, Kate and Handanagic, Senad and Crepaz, Nicole and Rosenberg, Eli S.},
  journal={Clinical Infectious Diseases},
  volume={76},
  number={1},
  pages={96--102},
  year={2023},
  publisher={Oxford University Press}
}

@article{stocks2020,
  title={Dynamic modeling of hepatitis {C} transmission among people who inject drugs},
  author={Stocks, Theresa and Martin, Leah J. and K{\"u}hlmann-Berenzon, Sharon and Britton, Tom},
  journal={Epidemics},
  volume={30},
  pages={100378},
  year={2020},
  publisher={Elsevier}
}

@article{diekmann1990,
  title={On the definition and the computation of the basic reproduction ratio {$\mathcal{R}_0$} in models for infectious diseases in heterogeneous populations},
  author={Diekmann, Odo and Heesterbeek, Johan A. P. and Metz, Johan Anton Jacob},
  journal={Journal of Mathematical Biology},
  volume={28},
  number={4},
  pages={365--382},
  year={1990},
  publisher={Springer}
}

@article{vandendriessche2002,
  title={Reproduction numbers and sub-threshold endemic equilibria for compartmental models of disease transmission},
  author={van den Driessche, P. and Watmough, James},
  journal={Mathematical Biosciences},
  volume={180},
  number={1--2},
  pages={29--48},
  year={2002},
  publisher={Elsevier}
}

@book{lenhart2007optimal,
  title={Optimal Control Applied to Biological Models},
  author={Lenhart, Suzanne and Workman, John T.},
  year={2007},
  publisher={Chapman and Hall/CRC}
}

@article{petzold1983automatic,
  title={Automatic selection of methods for solving stiff and nonstiff systems of ordinary differential equations},
  author={Petzold, Linda},
  journal={SIAM Journal on Scientific and Statistical Computing},
  volume={4},
  number={1},
  pages={136--148},
  year={1983},
  publisher={SIAM}
}

@article{marino2008methodology,
  title={A methodology for performing global uncertainty and sensitivity analysis in systems biology},
  author={Marino, Simeone and Hogue, Ian B. and Ray, Christian J. and Kirschner, Denise E.},
  journal={Journal of Theoretical Biology},
  volume={254},
  number={1},
  pages={178--196},
  year={2008},
  publisher={Elsevier}
}

@article{hofmeister2019estimating,
  title={Estimating prevalence of hepatitis {C} virus infection in the {United States}, 2013--2016},
  author={Hofmeister, Megan G. and Rosenthal, Elizabeth M. and Barker, Laurie K. and Rosenberg, Eli S. and Barranco, Meredith A. and Hall, Eric W. and Edlin, Brian R. and Mermin, Jonathan M. and Ward, John W. and Ryerson, A. Blythe},
  journal={Hepatology},
  volume={69},
  number={3},
  pages={1020--1031},
  year={2019},
  publisher={Wiley Online Library}
}

@article{naturaldeathrate,
  author  = {Ahmad, Farida B. and Cisewski, Jodi A. and Anderson, Robert N.},
  title   = {Mortality in the {United States} Provisional Data, 2023},
  journal = {MMWR Morbidity and Mortality Weekly Report},
  volume  = {73},
  number  = {31},
  pages   = {677--681},
  year    = {2024},
  doi     = {10.15585/mmwr.mm7331a1}
}

@misc{cdc2018hepatitis,
  author       = {{Centers for Disease Control and Prevention}\hspace{0pt}},
  title        = {Viral Hepatitis Surveillance Report -- {United States}, 2018},
  year         = {2020},
  note         = {https://archive.cdc.gov/www\_cdc\_gov/hepatitis/statistics/2018surveillance/HepC.htm}
}

@misc{cdc2019hepatitis,
  author       = {{Centers for Disease Control and Prevention}\hspace{0pt}\hspace{0pt}},
  title        = {Viral Hepatitis Surveillance Report -- {United States}, 2019},
  year         = {2021},
  note         = {https://www.cdc.gov/hepatitis/statistics/2019surveillance/HepC.htm}
}

@misc{cdc2020hepatitis,
  author       = {{Centers for Disease Control and Prevention}\hspace{0pt}\hspace{0pt}\hspace{0pt}},
  title        = {Viral Hepatitis Surveillance Report -- {United States}, 2020},
  year         = {2022},
  note         = {https://www.cdc.gov/hepatitis/statistics/2020surveillance/hepatitis-c/figure-3.1.htm}
}

@misc{cdc2021hepatitis,
  author       = {{Centers for Disease Control and Prevention}\hspace{0pt}\hspace{0pt}\hspace{0pt}\hspace{0pt}},
  title        = {Viral Hepatitis Surveillance Report -- {United States}, 2021},
  year         = {2023},
  note         = {https://www.cdc.gov/hepatitis-surveillance-2021/hepatitis-c/figure-3-1.html}
}

@article{leyva2020,
  author  = {Leyva, Yuridia and Page, Kimberly and Shiboski, Stephen and
             Hahn, Judith A. and Evans, Jennifer and Erhardt, Erik},
  title   = {Per-Contact Infectivity of Hepatitis {C} Virus Acquisition in
             Association With Receptive Needle Sharing Exposures in a
             Prospective Cohort of Young Adult People Who Inject Drugs in
             {San Francisco, California}},
  journal = {Open Forum Infectious Diseases},
  volume  = {7},
  number  = {4},
  pages   = {ofaa092},
  year    = {2020},
  doi     = {10.1093/ofid/ofaa092}
}

\section*{Declaration of competing interest}
The author declares no competing interests.

\section*{Acknowledgments}
The authors thank anonymous reviewers for their valuable suggestions. 

\section*{Funding}
This research did not receive any specific grant from funding agencies
in the public, commercial, or not-for-profit sectors.

\end{document}